\documentclass[a4paper,11pt]{article}
\pdfoutput=1 

\usepackage{jheppub} 
\usepackage{comment}
\usepackage{color}
\usepackage{mathtools}
\usepackage[T1]{fontenc} 
\usepackage{cancel}
\usepackage{cleveref}
\crefformat{section}{\S#2#1#3} 
\crefformat{subsection}{\S#2#1#3}
\crefformat{subsubsection}{\S#2#1#3}

\title{Soft Constraints on KMOC Formalism}

\author[a,b]{Yilber Fabian Bautista,}\emailAdd{ybautistachivata@perimeterinstitute.ca}
\author[c]{Alok Laddha}\emailAdd{aladdha@cmi.ac.in}

\affiliation[a]{Perimeter Institute for Theoretical Physics, 31 Caroline St N, Waterloo ON, N2L 2Y5, Canada}
\affiliation[b]{Department of Physics   and  Astronomy, York University, 4700 Keele St, Toronto ON, M3J 1P3, Canada}
\affiliation[c]{Chennai Mathematical Institute, H1, SIPCOT IT Park, Siruseri, Kelambakkam 603103, India}

\abstract{
In  this note, we investigate the implications of classical soft theorems for the formalism developed by Kosower, Maybee and O'Connell (KMOC) to derive classical observables in gauge theory and gravity from scattering amplitudes.  In particular, we show that the radiative electro-magnetic field at leading order in the soft expansion imposes an  infinite hierarchy of constraints on  the expectation value of the family of observables generated by \textit{monomials} of linear impulse. 
We perform an explicit check on these constraints at next to leading order (NLO) in the coupling and as a corollary show how up to NLO, soft radiation obtained from quantum amplitudes is consistent with the (leading) classical soft photon theorem. 

We also argue that in 4 dimensions the classical log soft theorem derived by Saha, Sahoo and Sen generates an infinite hierarchy of constraints on the expectation value of operators which are products of one angular momentum and an arbitrary number of linear momenta.}
\begin{document} 
\maketitle
\flushbottom

\section{Introduction}
\label{sec:intro}
Classical soft photon (graviton) theorems  \cite{ Laddha:2018rle,Laddha:2018myi, Sahoo:2018lxl,Laddha:2019yaj, Saha:2019tub}  are exact statements about  soft radiation emitted during a generic  electro-magnetic (gravitational) scattering process. As shown in the seminal works by Sahoo and Sen\cite{Sahoo:2018lxl}, Saha, Sahoo and Sen \cite{Saha:2019tub}, and Sahoo \cite{Sahoo:2020ryf}, in four dimensions if we expand the electro-magnetic (or gravitational) radiative field in the frequency of the emitted radiation, then the following terms in the expansion have a universal analytic form independent of the details of the scattering dynamics or even spins of the scattering particles
\begin{flalign}\label{one}
A_{\mu}(\omega, \hat{n})\, =\, \frac{1}{\omega}\, A_{\mu}^{\omega^{-1}}(\hat{n})\, +\, \sum_{I=1}^{2}\,  \omega^{I} (\ln\omega)^{I+1}\, A_{\mu}^{\ln^{I+1}}(\hat{n})\, +\, \cdots\,.
\end{flalign}
Here $\hat{n}$ is a unit vector pointing towards the direction of observation, 
and  $\cdots$ indicate sub-sub-leading terms in the soft expansion. It was conjectured in \cite{Sahoo:2020ryf} that even among the sub$^{n}$-leading  terms the coefficients of $\omega^{n}\, \ln^{n+1}\omega\  n\, \geq\, 3$ are universal while other terms in the soft expansion are non-universal and depend on the details of the dynamics. In \cite{Ghosh:2021hsk}, first such non-universal soft factor proportional to $\omega\ln\omega$ was computed and was shown to depend on the spin of the scattering particles. 

Each coefficient in the above expansion is a function of incoming and outgoing momenta and charges of the scattering particles.  For example, the leading coefficient $A_{\mu}^{\omega^{-1}}(\hat{n})$ is simply the Weinberg soft photon factor \cite{PhysRev.140.B516} 
\begin{flalign}
A_{\mu}^{\omega^{-1}}(\hat{n})\, =\, \sum_{i=1}^{N_{in}}\, Q_{i}\, \frac{1}{p_{i} \cdot n}\, p_{i}^{\mu}\, -\, \sum_{i=1}^{N_{out}}\, Q_{i}\, \frac{1}{p_{i} \cdot n}\, p_{i}^{\mu}\,,
\end{flalign} 
where  $\{(Q_{1}, p_{1}),\, \cdots,\, (Q_{i}, p_{i})\, \}$ is the collection of  the charges and momenta of scattering particles, and   $n^{\mu}\, =\, (1, \hat{n})$.  Although the exact expressions for sub-leading and higher order log soft factors in eqn.(\ref{one}) are more complicated, they are all functions of asymptotic data, namely charges and momenta of scattering particles.  

From the perspective of scattering dynamics, these theorems are rather non-trivial as they are non perturbative in the coupling.\footnote{In an interesting recent work \cite{Freidel:2021qpz}, an attempt has been made to analyse the infinite set of constraints on the gravitational dynamics from  asymptotic symmetries which are in turn related to classical soft theorems.} If we consider a class of scattering processes which can be analysed perturbatively (such as large impact parameter scattering so that $N_{in} = N_{out} = N$) then every outgoing momentum $p_{i\, +}^{\mu}$,  admits a perturbative expansion in terms of the incoming momentum $ p_{i\, -}^{\mu} $, and the impulse acquired by the particles at every order in a perturbative expansion \footnote{We note that from the classical perspective, this expansion is convergent as final momenta are well defined.} 
\begin{flalign}\label{eq:expansion_momenta}
p_{i\, +}^{\mu}\, =\, p_{i\, -}^{\mu} + \sum_{L=0}\, g^{2(L+1)}\, (\triangle p^{(L)})_{i}^{\mu}\,.
\end{flalign}
Here $g$ is the coupling\footnote{We have use $g$ instead of $e$, the charge of the electron in order to avoid confusion with the exponential function.} and  $(\triangle p^{(L)})^{\mu}$ is the linear impulse evaluated at   $L$-th order in the perturbation theory ($L=0$ being the leading order (LO) impulse). We thus see when expanded in the coupling, the Weinberg soft photon factor has a rather intricate structure
\begin{flalign}\label{two} 
A_{\mu}^{\omega^{-1}}(\hat{n})\, =\, \sum_{s=1}^{N}\, g\,Q_{s} \sum_{n=0} g^{2(n+1)}\, \sum_{i=0}^{n}\, V^{\mu}_{s\, \alpha_{1}\, {\cdots}\, \alpha_{i+1}}\ \smashoperator{\sum_{\substack{L_{1}+\, \cdots+\, L_{i+1}=0 \\ (L_{1} + L_{i+1}) + (i+1)\, = n}}^{n +1 - i}}\ (\, (\triangle p^{(L_{1})})^{\alpha_{1}}\, {\cdots}\, (\triangle p^{(L_{i+1})})^{\alpha_{i+1}}\, )\,,
\end{flalign}
where the sum $\sum_{L_{1},\, \cdots,\, L_{i+1}=0\, \vert\, (L_{1} + L_{i+1}) + (i+1)\, = n}^{n +1 - i}$ is over products of impulses at $L_{1},\, \cdots,\, L_{i+1}$ orders in the coupling respectively. In the above equation we have defined a tensor 
\begin{flalign}\label{eq:projector_general_s}
V^{\mu}_{s\, \alpha_{1}\, \cdots\, \alpha_{i+1}}\, =\, \textcolor{black}{(-1)^i}\, \Big[\, \frac{ (i+1)\, !\,}{(p_{s} \cdot k)^{i+1}} \delta^{\mu}_{(\alpha_{1}}\, k_{\alpha_{2}}\, \cdots\, k_{\alpha_{i+1})}\, -\,  \frac{1}{(p_{s} \cdot k)^{i+2}}\, p_{s}^{\mu}\, k_{\alpha_{1}}\, \cdots\, k_{\alpha_{i+1}}\,\Big]\,.
\end{flalign}
Classical Soft photon (or graviton) theorem are independent of the details of the hard scattering and are applicable to perturbative scattering at finite impact parameter as well as collisions.  However, to prove the classical soft theorems via perturbative analysis (even in the case where hard scattering can be treated perturbatively) is a highly non-trivial task as one has to resum the perturbation series. But the discussion above demonstrates that, 
due to their universality, classical soft theorems can serve as  powerful tool for any method which computes electro-magnetic (or gravitational) radiation using (perturbative) scattering amplitudes. For one, it can serve as a strong diagnostic for the perturbative results of radiation  and when used in conjunction with the perturbative results (such as analytic expressions for impulse in the Post Minkowskian (PM) expansion), it can produce interesting insights such as providing analytical formulae for classical radiation in terms of incoming kinematic data order by order in perturbation theory. 

One such methods aforementioned,  was  developed by Kosower, Maybee and O'Connell (KMOC) \cite{Kosower:2018adc}, whose formalism allows to compute classical electromagnetic (gravitational) observables from the classical limit of quantum scattering amplitudes.  These observables include the linear and angular momentum impulse -- including spin -- experienced by a scattering particle, as well as radiative field emitted in a classical scattering \cite{Kosower:2018adc,Maybee:2019jus,Guevara:2019fsj,Cristofoli:2021vyo,Aoude:2021oqj,Herrmann:2021lqe}. 
In this paper we initiate a study of the implications of classical soft theorems for KMOC formalism. As we show, consistency with the leading classical soft theorem imposes an infinite hierarchy of  constraints on KMOC observables. In order to state these constraints we introduce following conventions. 

The scattering process we consider is a $2\, \rightarrow\, 2$ scattering process in which two incoming charged particles with momenta $p_{1}, p_{2}$ and charges $Q_{1},\, Q_{2}$ scatter via electro-magnetic interactions as well as any other higher derivative interaction which is long range such that the KMOC formalism applies to this scattering. As the classical soft theorems are universal and independent of the details of the scattering, \textcolor{black}{the low frequency classical radiative field} obtained via KMOC formalism should generate the soft factors for \emph{any} perturbative amplitude involving charged particles in the external states and a photon. \textcolor{black}{It is particularly  remarkable how the classical field is   controlled by  \textit{single} photon emission amplitudes, while the classical field should be composed from many photon. In \cite{Cristofoli:2021jas}, it was shown such single photon emission amplitude parametrize the high photon occupation number as expected for a classical field.} 

To each of the two massive particles we can associate certain classical observables defined as follows:\\
(1) Let  $V^{\mu}_{\alpha_{1}\, \cdots\, \alpha_{i}}$ be the  projection operator \eqref{eq:projector_general_s}, associated to particle $1$, that is
\begin{flalign}\label{eq:soft_projector_1}
V^{\mu}_{\alpha_{1}\, \cdots\, \alpha_{i}}\, =\, (-1)^{i+1}\, \Big[\, \frac{i!}{(p_{1} \cdot k - i\epsilon)^{i}}\,\delta^{\mu}_{(\alpha_{1}}\, k_{\alpha_{2}}\, \cdots\, k_{\alpha_{i})}\, - \frac{1}{(p_{1} \cdot k + i\epsilon)^{i+1}}\,p_{1}^{\mu}\, k_{\alpha_{1}}\, \cdots\, k_{\alpha_{i}}\, \Big]\,.
\end{flalign}
(2) Now consider certain  \textit{moments} of the exchange momenta 

\begin{flalign}\label{eq:moments}
\begin{array}{lll}
{\cal T}^{(n)\,\alpha_{1}\, \cdots\, \alpha_{i}}\, :=
 \,\hbar^{\frac{3}{2}}\Big[ i\int\, \hat{d}\mu_q\, q^{\alpha_{1}}\, \cdots\,  q^{\alpha_{i}}\,e^{-iq{\cdot}b} \, {\cal A}^{(n)}(p_{1},p_2\, \rightarrow\, p_{1} - q, p_{2} + q)\\
\hspace*{3.5cm}+\, \sum_{X=0}^{n-1}\,\int\, \prod_{m=0}^X d\Phi(r_m) \,  \hat{d}\mu_q\, \hat{d}\mu_{w,X}  w^{\alpha_{1}}\, \cdots\, w^{\alpha_{i}}\,e^{-iq{\cdot}b} \\
\hspace{4.5cm} \times \sum_{a_{1}=0}^{n-1-X} {\cal A}^{(a_{1}) }(p_{1},p_2\, \rightarrow\, p_{1} - w, p_{2} + w, \, r_{X})\,\\
\hspace*{4.5cm}\times {\cal A}^{(n-a_{1} - X-1)\star}(p_{1} - w,p_2+w, \, r_{X}\, \rightarrow\, p_{1} - q,p_2+q)\Big]\, \,,
\end{array}
\end{flalign}
where $b$ is the impact parameter in the $2\to2$ scattering process. In the above equation we have introduced the following notations which will be used throughout the paper. 
\begin{itemize}
\item ${\cal A}^{(a)}( p_{1}^{i},\, p_{2}^{i}\, \rightarrow\, p_{1}^{f}, p_{2}^{f})$ is the reduced amplitude (where the momentum conserving $\delta$ function has been factored out) for a   4 point scattering at $a$-th loop order, and analogously for the other amplitudes with additional momentum labels. 
\item The  integral measure $\hat{d}\mu_q$ is defined via
\begin{equation}\label{eq:measure}
    \hat{d}\mu_q = \hat{d}^4q\hat{\delta}(2p_1{\cdot}q-q^2)\hat{\delta}(2p_2{\cdot}q+q^2),
\end{equation}
where $\hat{d}^4q =\frac{d^4q}{ (2\pi )^4}$, and the 
 hat on $\delta$-function indicates it is defined as
\begin{flalign}
\hat{\delta}(x)\, =\, -i\, [\, \frac{1}{x-i\epsilon}\, - \frac{1}{x + i\epsilon}\, ]\,,
\end{flalign}
and analogous  for $\hat{d}\mu_{w,X}$,
\begin{equation}\label{eq:measurex}
    \hat{d}\mu_{w,X} = \hat{d}^4q\hat{\delta}(2p_1{\cdot}w-w^2)\hat{\delta}(2p_2{\cdot}(w+r_X)+(w+r_X)^2)
\end{equation}
\item The sum over $X$ is a sum over number of intermediate photons with momenta $\{ r_{1}\, \cdots,\, r_{X}\, \}$. Even though integration over the momentum space of these photons is indicated explicitly by $d\Phi(r_m)=\hat{d}^4r_m\hat{\delta}^{(+)}(r_m^2)$, we assume that the sum over $X$ includes the sum over intermediate helicity states. Notice also these  conventions for intermediate photons momenta labels assume $r_0 = 0$ and $\int d\Phi(r_0)\to 1 $, which effectively recovers the  contribution from the conservative sector. 
\item It is understood that for $n=0$, the second term in \eqref{eq:moments} vanishes. 
\end{itemize}
As we will show in \cref{sec:weinberg_kernel}, consistency of KMOC with the  classical leading soft photon theorem \cite{Saha:2019tub}, implies that at $n$-order in perturbative expansion  we have the following identities
\begin{itemize}
    \item $\forall\, n>0$ and $\forall\, 1\, \leq\, i\, \leq\, n$ :
\end{itemize}
\begin{equation}\label{hier1}
\begin{split}
 \lim_{\hbar \rightarrow\, 0}\, \hbar^{m}\, V^{\mu}_{\alpha_{1}\, \cdots\, \alpha_{i}}\, {\cal T}^{(n)\,\alpha_{1} \cdots \alpha_{i}}\, =\, 0\, \forall\, m\, \in\, \{1,\cdots,n+ 1-i\}\, ,
  \end{split}
\end{equation}
\begin{itemize}
    \item and  
$\forall\, n$ and $\forall\, 1\, \leq\, i\, \leq\, n\, +\, 1$ :
\end{itemize} 
\begin{equation}\label{hier2}
    \lim_{\hbar\, \rightarrow\, 0}\, V^{\mu}_{\alpha_{1}\,{\cdots}\, \alpha_{i}}  {\cal T}^{\alpha_{1}\, \cdots\, \alpha_{i}}_{(n)}\, = g^{2(n+1)} \, V^{\mu}_{\alpha_{1}\, \cdots\, \alpha_{i}}\, \smashoperator{\sum_{L_{1}+\, {\cdots}+\, L_{i-1} =0}^{n}}\, (\triangle p_{1}^{L_{1}})^{\alpha_{1}}\, {\cdots}\, (\triangle p_{1}^{n {-} (L_{1} {+} \cdots {+} L_{i-1})})^{\alpha_{i}}  \,.
\end{equation}
\textcolor{black}{Together, eqns. (\ref{hier1}, \ref{hier2}) are an infinity of constraints that the classical soft theorem impose on the  moments defined in the KMOC formalism. In particular, the first set of constraints given in  eqn.(\ref{hier1}) ensure that classical limit of soft radiative field is smooth and the second set imply that this smooth limit precisely equals the leading classical soft factor \emph{at all orders} in perturbative expansion.}
These constraints were shown to be satisfied at tree-level in the  earlier work of \cite{Bautista:2019tdr,Manu:2020zxl}, where at LO in the coupling,  leading and sub-leading classical soft photon theorem was derived from KMOC formula, which we will review in \cref{sec:tree-level-radiation} for the leading soft result. 

\textcolor{black}{Analogous constraints follow for the subleading soft factor, and we briefly comment on those in \cref{sec:subleading_soft_sonstraints}.  In particular as we argue, the classical log soft theorem implies that  the classical limit of  certain moments composed of linear and angular momentum operators are constrained by the log soft factor at all orders in perturbation theory.  We believe that these constraints along with similar ``higher-order constraints" that would be generated from the universality of sub-subleading ($\omega\ln\omega^{2}$) soft factor have a potential to generate a template that can directly express classical limit of highly intricate quantum operators in terms of known conservative quantities like the linear impulse.}

Let us here also stress that although in this note we are concerned mostly with 4-dimensional  electromagnetic radiation , the higher dimensional generalization of these constraints, as well as the equivalent formulation for gravitational radiation, should follow in a similar manner.

This paper is organized as follows: In \cref{sec:classical_review} we  review perturbative results for classical soft photon theorem at leading and subleading orders in the soft expansion. 
In \cref{sec:review_KMOC} we provide a short review of the KMOC formalism in the context of radiation. We then move to  the
derivation of identities \eqref{hier1} and \eqref{hier2} in \cref{sec:weinberg_kernel}. In section \cref{sec:subleading_soft_sonstraints} we argue that analogous constraints on  KMOC follow from the sub-leading soft photon theorem. 
In \cref{sec:lo-soft_costraitns} 
 we  show how the KMOC formula indeed satisfies these constraints at leading (\cref{sec:tree-level-radiation}) and next to leading (\cref{sec:1-loop_radiation}) order in the coupling, by working with amplitudes in scalar QED.  
That is, contribution of the tree-level and one loop amplitudes to  the $\frac{1}{\omega}$ coefficient of the radiative field indeed matches with eqn. (\ref{two}). \textcolor{black}{ The checked of the constraints \cref{sec:lo-soft_costraitns}  at NLO is analog to the cancellation of the superclassical terms for the NLO electromagnetic impulse studied in  \cite{Kosower:2018adc}, present in the same way  in the eikonal approach, see for instance \cite{DiVecchia:2021bdo}. } 
Finally, in \cref{sec:discusion} we conclude with a discussion. 
 In appendix \ref{app:radiation_NLO} we provide some computational details for \cref{sec:1-loop_radiation}, whereas in Appendix \ref{app:NLO-IMPULSE} we provide a explicit derivation of the electromagnetic impulse at NLO in the perturbative expansion directly from the KMOC formalism, and recovering the result of \cite{Saketh:2021sri}.

\section{Soft Radiation in Classical Scattering}\label{sec:classical_review}

In this section, we analyse the classical soft photon factor up to sub-leading order in soft expansion and NLO in the coupling in terms of explicit expressions for the linear impulse. We simply use the results in \cite{Saketh:2021sri}, in conjunction with classical soft theorem to write the radiative field at the desired order.  That is, we compute $A_{\mu}^{\omega^{-1}}(\hat{n})$ and $A_{\mu}^{\ln}(\hat{n})$ to NLO in the coupling in a classical scattering involving two charged particles with masses $m_{1}, m_{2}$ which are interacting only via electro-magnetic interactions. 

Although our primary focus in this paper from the perspective of KMOC is on the $\frac{1}{\omega}$ soft factor, we also give an explicit verification of the sub-leading ($\ln\omega$) soft photon theorem from explicit computation.  That is, we derive the logarithmically divergent (in asymptotic time) contribution to the angular momentum impulse and substitute it in the ``formal sub-leading soft factor" expression (which is obtained by substituting classical angular momentum in the angular momentum operators in Cachazo-Strominger \cite{Cachazo:2014fwa} soft factor $\frac{J^{\mu\nu}k_{\nu}}{p \cdot k}$)  to show that the result equals the log soft photon factor at NLO in the coupling \cite{Saha:2019tub}. 

\subsection{Leading soft factor}

\textit{Leading order  radiation}:

Let   $p_{i}\, \vert\, i = 1,2$ be the  momenta for in incoming massive particles, moving in the asymptotic free trajectories $b^\mu+v^\mu\tau$ in the far pass. If we denote the null vector $(1, \hat{n})$ as $n^{\mu}$ we can write the leading soft factor at tree level from formulas \eqref{two} and \eqref{eq:projector_general_s}, that is

\begin{equation}\label{eq:leading-soft-tree}
    A^{(0)\,\mu}_{\omega^{-1}}(\hat{n}) = g^3\sum_{i=1}^{2}Q_{i}\left[\frac{\Delta p_{i}^{(0)\,\mu}}{p_{i}{\cdot}n}-\frac{\Delta p_{i}^{(0)\,\mu}\cdot n}{(p_{i}{\cdot}n)^{2}}p_{i}^{\mu}\right]\,,
\end{equation}
where the leading order linear impulse has the well known form  \begin{equation}\label{eq:LO_Impulse_explicit_intr}
   \Delta p^{(0)\,\mu}_{1}= -\Delta p^{(0)\,\mu}_{2}=-\frac{Q_{1}Q_{2}\,p_{1}{\cdot}p_{2}}{2\pi\sqrt{\mathcal{D}}}\frac{b^{\mu}}{b^{2}},\,\,\,\,\,b^{2}=-\vec{b}^{2},
\end{equation}
with the Jacobian factor
\begin{equation}\label{eq:jacobian}
    \mathcal{D} = \left(p_{1}{\cdot}p_{2}\right)^{2}-m_{1}^{2}m_{2}^{2}.
\end{equation}

 This  simple examples shows explicitly how the  radiated field to leading order in the soft expansion  is determined only from asymptotic data, and in  particular for perturbation theory, from only incoming data since the outgoing momenta are determined by the perturbative expansion \eqref{eq:expansion_momenta}, which we have truncated at leading order in the coupling. Let us finally mention the leading order radiated field \eqref{eq:leading-soft-tree} encodes the so called linear memory effect of the scattering process \cite{Strominger:2017zoo}, which has direct analog in the gravitational case as first derived by Braginsky and Thorne \cite{Braginsky1987}.
 \\\\
\textit{Sub-Leading order radiation}

At NLO, the radiated field has a more interesting form, since as indicated in  \eqref{two} and \eqref{eq:projector_general_s}, both, the leading and subleading impulse enter into the field. Indeed, it explicitly reads 
\begin{equation}\label{eq:soft-one-loop}
    A_{\omega^{-1}}^{(1)\,\mu}(\hat{n}) =g^5 \sum_{i=1}^{2}Q_{i}\left[\frac{\Delta p_{i}^{(1)\,\mu}}{p_{i}{\cdot}n}-\frac{\Delta p_{i}^{(1)}\cdot n}{(p_{i}{\cdot}n)^{2}}p_{i}^{\mu}-\frac{\Delta p_{i}^{(0)}\cdot n}{(p_{i}{\cdot}n)^{2}}\Delta p_{i}^{(0)\,\mu}+\frac{(\Delta p_{i}^{(0)}\cdot n)^{2}}{(p_{i}{\cdot}n)^{3}}p_{i}^{\mu}\right]\,.
\end{equation}
At this order, it is still true that $\Delta p_{1}^{(1)\,\mu}=-\Delta p_{2}^{(1)\,\mu}$, with the NLO impulse given  explicitly  by \cite{Saketh:2021sri} (upon setting $g\to\frac{g}{\sqrt{4\pi}}$),

\begin{equation}\label{eq:NLO-impulse-justin}
\Delta p_{1}^{(1)\,\mu}=-\frac{g^{4}}{32\pi^2|b|^{3}}\frac{(Q_{1}Q_{2})^2}{\mathcal{D}}\left[\pi\sqrt{\mathcal{D}}(m_{1}+m_{2}) b^{\mu}+4\frac{(p_{1}{\cdot}p_{2})^2(p_{1}{+}p_{2})^{2}|b|}{\mathcal{D}}p^{\mu}\right]\,,
\end{equation}
where we have used 
\begin{equation}\label{eq:p-def_justin}
    p^\mu = \frac{m_1m_2}{(p_1+p_2)^2}\left[\left(\frac{m_2}{m_1}+ \frac{p_1\cdot p_2}{m_1 m_2}\right)p_1^\mu -  \left(\frac{m_1}{m_2}+ \frac{p_1\cdot p_2}{m_1 m_2}\right)p_2^\mu\right]\,.
\end{equation}
In Appendix \ref{app:NLO-IMPULSE} we show how to recover this result from KMOC formalism. 
\subsection{Sub-leading soft factor} \label{sec:subleading_soft_factor_classical}
In \cite{Laddha:2018myi} it was argued  that the classical form of the  Cachazo-Strominger soft factor is in fact of the order $\ln\omega$ and not $\omega^{0}$ in the soft expansion. This is due to the fact that  when scattering particles are subjected to  long range electro-magnetic (or gravitational)  forces, the correction to the free trajectory $b^{\mu}\, +\, v^{\mu}\, \tau$ is not a power series in $\frac{1}{\tau}$ but include logarithmic tails such as $\ln \tau$. As a result of this long range effect, the orbital angular momentum of the scattering particles is dominated by terms proportional to $\ln \tau$ at late times. Intuitively we expect late times to correspond to soft frequencies and hence in the frequency space, the asymptotic orbital angular momentum is expected to be dominated by terms proportional to $\ln\omega$. Thus, the Cachazo-Strominger soft factor which naively is of the order $\omega^{0}$, in fact generates the so-called log soft factor. This argument was turned into a theorem in a series of remarkable papers by Saha, Sahoo and Sen. \cite{Sahoo:2018lxl, Saha:2019tub}.  
More in detail, in the case of $2\, \rightarrow\, 2$ scattering, we can compute the  difference in angular momenta at late and early times  for both the scattering particles. As we show below, this difference indeed has the form
\begin{flalign}
J_{\mu\nu}^{+}\, -\, J_{\mu\nu}^{-}\, =\, \ln\tau\, (\triangle J)_{\mu\nu}^{(\ln)}\, +\, (\triangle J)_{\mu\nu}^{(0)}\, ,
\end{flalign}
where $\ln\tau$ is understood as an logarithmic divergent term as $\tau\, \rightarrow\, \infty$.\footnote{As was argued in \cite{Laddha:2018myi} frequency space we expect $\ln\tau$ to be replaced by $\ln\omega^{-1}$ for soft frequency $\omega$. This divergence shows that soft expansion in $\omega$ breaks down in Four dimensions.} 
We can now use the above argument to ``guess" the log soft photon factor.
\begin{flalign}\label{logsoftc}
A_{\mu}^{\ln}(\hat{n})\, =\, -\, \sum_{i=1}^{2}\, Q_{i}\,  \frac{1}{p_{i} \cdot n}\, \Big[\, (\triangle J)_{\mu\nu}^{\ln (0)}\, n^{\nu}\, +\, \Big( (\triangle J)_{\mu\nu}^{(\ln)\, (1)}\, n^{\nu}\, -\, \frac{1}{p_{i} \cdot n} (\triangle p_{i}^{(0)} \cdot n)\, (\triangle J)_{\mu\nu}^{\ln (0)}n^{\nu} \Big)\, \Big]\,,
\end{flalign}
where $(\triangle J)_{\mu\nu}^{\ln (i)}$ is the logarithmically divergent term in angular momentum impulse at N$^{i}$LO order in coupling. 

In this part of the note we use the  results provided in \cite{Saketh:2021sri}, to compute $\triangle J_{\mu\nu}^{\ln}$ at leading and NLO in the coupling.  Along with expression for $(\triangle p^{(0)})^{\mu}$, one can then obtain the log soft factor up to $O(g^{5})$ in the coupling. 

Let us see, in \cite{Saketh:2021sri}, the authors compute the deflection to the free trajectory up to N$^{2}$LO order in the perturbative expansion. That is, we expand the trajectory for the $i$-th particle with charge $Q_{i}$ and initial momentum $p_{i -}$ as
\begin{flalign}
x_{i}^{\mu}(\tau)\, =\, b_{i}^{\mu}\, +\, \frac{1}{m_{i}}\, p_{i -}^{\mu}\tau\, +\, \sum_{n=0}\, e^{2(n+1)}\,  (\delta^{n}x_{i})^{\mu}(\tau)\,,
\end{flalign}
where $(\delta^{n}x_{i})^{\mu}$ is the deflection at N$^{n}$LO order. At every order in the perturbative expansion the deflected trajectory accelerates as $O(\frac{1}{\tau^{2}})$ at late future times $\tau_{+}$. 
\begin{flalign}
(\delta^{n}x_{i})^{\mu}(\tau_{+})\, =\, (\delta^{n}x_{i})^{\mu}_{1}\ \tau_{+} +\, (\delta^{n}x_{i})^{\mu}_{\ln}\, \ln\tau_{+} + O(\tau_+^{0})\,.
\end{flalign}
The $\ln\tau_{+}$ co-efficient of the angular momentum impulse can now be computed as
\begin{flalign}
\begin{array}{lll}
(\triangle J_{i}^{\ln (0)})\, =\, (\delta^{0}x_{i})_{\ln}\, \wedge\, p_{i -}\,,\\
(\triangle J_{i}^{\ln (1)})\, =\, (\delta^{1}x_{i})_{\ln}\, \wedge\, p_{i -}\, +\, (\delta^{0}x_{i})_{\ln}\, \wedge\, (\triangle p_{i})^{(0)}\,.
\end{array}
\end{flalign}
Both $\delta^{0}x_{\ln},\, \delta^{1}x_{\ln}$ can be read off from expressions given by equations $(2.24)$ and $(\text{A}25)$ in  \cite{Saketh:2021sri}.  We focus on the radiative field sourced by the first particle with initial momentum $p_{1}^{\mu}$. \footnote{We suppress the label $-$ to indicate the initial momenta as it will not be required in the rest of the section}
\begin{flalign}
\begin{array}{lll}
(\delta^{0}x_{1})^{\mu}_{\ln}\, =\, Q_{1}\, Q_{2}\, \frac{1}{m_{1} (\gamma v)^{3}}\, (-\gamma v_{1}^{\mu} + v_{2}^{\mu} )\\
(\delta^{1}x_{1})^{\mu}_{\ln} = \frac{Q_{1}^{2} Q_{2}^{2}}{m_{1}^{2}}\, \Big[\, - \frac{\gamma^{3} v \hat{b}^{\mu}}{\vert b \vert (\gamma v)^{5}}\, - \frac{m_{1}}{m_{2}}\, \frac{\gamma^{2}v\, \hat{b}^{\mu}}{\vert b\vert (\gamma v)^{5}}\, \Big]\,,\\
\hspace*{0.6in}=\,- \frac{Q_{1}^{2} Q_{2}^{2}}{m_{1}^{2}}\, \frac{\hat{b}^{\mu}}{\vert b\vert (\gamma v^{2})^{2}} \Big[\, 1\, + \frac{m_{1}}{m_{2}\gamma}\, \Big]\,.
\end{array}
\end{flalign}
The leading order perturbation of the final momenta from the initial momenta can be easily computed and is given by \footnote{This is the same impulse given in \eqref{eq:LO_Impulse_explicit_intr}, but here we have used the  normalization followed  in \cite{Saketh:2021sri}.}, 
\begin{flalign}
(\delta^{0}p_{1})^{\mu}\, =\, Q_{1}Q_{2} \frac{1}{v \vert b\vert}\, \hat{b}^{\mu}\,,
\end{flalign} 
where $\gamma\, =\, v_{1}\, \cdot v_{2}\, =\, \sqrt{\frac{1}{1-v^{2}}}$. One can write the Lorentz factors in a more familiar form as, 
\begin{flalign}
\begin{array}{lll}
\frac{1}{\gamma v}\, =\, \frac{m_{1}m_{2}}{{\cal \sqrt{D}}}\,.
\end{array}
\end{flalign}
Next we can  use these formulae for logarithmic deflections in the far future to compute the $\ln\tau_{+}$ contribution to angular momentum impulse at LO and NLO in the coupling
\begin{equation}
    \triangle J^{\ln (0)}\, =\, 
Q_{1}\, Q_{2}\, \frac{(m_{1}m_{2})^{2}}{{\cal D}^{\frac{3}{2}}}\, p_{2}\, \wedge\, p_{1}\,,
\end{equation}
and 
\begin{equation}
   \triangle J^{\ln\, (1)}\, =\, -2\frac{(Q_1 Q_2\,m_1 m_2)^2 p_1{\cdot}p_2}{b^2\mathcal{D}^{2}}b\wedge(p_1+p_2)\,, 
\end{equation}
respectively. On substituting these expressions in eqn.(\ref{logsoftc}), one can explicitly verify that up to NLO,  the result precisely matches with the classical soft factor derived by Saha, Sahoo and Sen in \cite{Saha:2019tub}.

\section{ KMOC Formalism in a Nutshell
 }\label{sec:review_KMOC}

In the KMOC formalism  \cite{Kosower:2018adc}, the expectation value for the change of a  classical  observable $\langle  \mathcal{O} \rangle $\footnote{We use $\langle...\rangle$ to imply that the classical limit for the given observable is taken. } during a scattering process, is computed from the scattering matrix through the formula 
\begin{equation}\label{eq:kmoc_formula}
    \langle  \mathcal{O} \rangle = \lim_{\hbar\to 0}\hbar^{\beta_\mathcal{O}}\left[_{\rm{in}}\langle\psi|S^\dagger \hat{\mathcal{O}}S|\Psi\rangle_{\rm{in}} -\, _{\rm{in}}\langle\psi| \hat{\mathcal{O}}|\Psi\rangle_{\rm{in}} \right]\,,
\end{equation}
where  $-\beta_{\mathcal{O}}$ is the power of the LO-piece in the $\hbar$-expansion of the quantities inside the square brackets,  which depends on the specific observable, as well as on the theory considered. Then, the factor of $\hbar^{\beta_\mathcal{O}}$ in this formula  ensures  $ \langle  \mathcal{O} \rangle\sim \hbar^0$, i.e. classical scaling. For instance,   for the radiated photon field, we have $\beta_{\mathcal{O}}=\frac{3}{2}$.

In formula \eqref{eq:kmoc_formula},  $|\Psi\rangle_{in}$ corresponds to  the multi-particle initial state, describing particles as wave functions with well-defined positions and momenta. In this note we are interested in a $2\to 3$ scattering process, then, we take  the 2-particles initial state to be given by 
\begin{equation}
    |\Psi\rangle_{in} =\int \prod_i \big[\hat{d}^4p_i
   \hat{\delta}^{(+)}(p_i^2-m_i^2)\phi_i(p_i)e^{ib_i{\cdot}p_i/\hbar}\big] |p_1p_2\rangle
\end{equation}
where we have employed the  notation of the original reference \cite{Kosower:2018adc}, however, unlike for the original work, and to be more general,  we have move to a frame where both particles are displaced by the positions $b_i$, whose difference, $b_2-b_1=b$, corresponds then to the impact parameters, which is the  distance of closest approach  between the massive particles during the scattering process. The next task is to relate the observable to the scattering amplitude  
employing  the usual representation of the scattering matrix in terms of the transition matrix  $S=1+iT$, followed by the classical limit, the latter of which is achieved by the usual $\hbar$-rescaling of the  coupling constant $g\to \hbar^{-1/2} g$, the 
 external massless momenta $k_i\to \hbar k_i$, the momenta miss-match $q_i=p_i'-p_i\to \hbar q_i$, where $p_i'$ is the momentum label for the conjugate states,   the loop momenta $l_i\to \hbar l_i$, and the position displacements $b_i\to \hbar^{-1}b$ (we point to  the original reference \cite{Kosower:2018adc}, for a detailed analysis of the classical limit).  
At this stage,  the  explicit dependence on the wave functions drops  away, leaving us with the desired expression for the classical observable. 

In this note, the observable of  interest  is the radiative electro-magnetic field $A^{(n)\,\mu}(k)$ \footnote{The radiative field is an observable as it is defined at null infinity where (small) gauge transformations vanish.}
where $k$ is the momentum of the electro-magnetic wave,  at  $\rm{N}^n\rm{LO}$ order in the perturbative expansion.    Hence, it is understood  that the $n=0$ term  represents the tree level contribution, whereas for $n>0$ we pick up loop corrections. It is a straightforward task to show that in terms of the scattering   amplitude, the radiated field can be computed from the sum of two terms 
\begin{equation}
 \langle A^{(n)\,\mu} (k)\rangle = \mathcal{R}^{(n)\,\mu}(k)+\mathcal{C}^{(n)\,\mu}(k),
\end{equation}
which   have the explicit recursive form 
\begin{equation}\label{eq:r-kernel_}
    \mathcal{R}^{(n)\,\mu}(k)=i\lim_{\hbar\to0}\hbar^{\frac{3}{2}}\int\prod_{i=1}^{2}\hat{d}^{4}q_{i}\hat{\delta}(2p_{i}{\cdot}q_{i}-q_{i}^{2})e^{ib_{i}{\cdot}q_{i}}\hat{\delta}^{4}(q_{1}{+}q_{2}-k)\mathcal{A}^{(n)\,\mu}\left(p_{1},p_{2}{\rightarrow}p_{1}{-}q_{1},p_{2}{-}q_{2},k\right)\,,
\end{equation}
and 
\textcolor{black}{
\begin{equation}\label{c-k}
\begin{split}\mathcal{C}^{(n)\,\mu}(k) & ={\color{black}}\lim_{\hbar\to0}\hbar^{\frac{3}{2}}\sum_{X=0}^{n-1}\int\prod_{m=0}^{X}d\Phi(r_{m})\prod_{i=1}^{2}\hat{d}^{4}w_{i}\hat{d}^{4}q_{i}\hat{\delta}(2p_{i}{\cdot}q_{i}{-}q_{i}^{2})\hat{\delta}(2p_{i}{\cdot}w_{i}{-}w_{i}^{2})e^{ib_{i}{\cdot}q_{i}}\\
 & \hspace{2cm}\,\,\,\,\,\,\,\,\,\,\,\,\times\hat{\delta}^{4}(w_{1}{+}w_{2}{+}r_{X}{-}k)\hat{\delta}^{4}(w_{1}{+}w_{2}{+}r_{X}{+}q_{1}{+}q_{2})\\
 & \hspace{2cm}\,\,\,\,\,\,\,\,\,\,\,\,\,\times\sum_{a_{1}=0}^{n-1-X}\mathcal{A}^{(a_{1})\,\mu}(p_{1},p_{2}{\rightarrow}p_{1}{-}w_{1},p_{2}{-}w_{2},r_{X},k)\\
 & \hspace{2cm}\,\,\,\,\,\,\,\,\,\,\,\,\,\times\mathcal{A}^{(n-a_{1}-X-1)\star}(p_{1}{-}w_{1},p_{2}{-}w_{2},r_{X}{\rightarrow}p_{1}-q_{1},p_{2}-q_{2})\,,
\end{split}
\end{equation}}
where the $\star$ in one  of the amplitude indicates complex conjugation.
We refer to the $\mathcal{C}^{(n)\,\mu}(k)$ term as the \textit{cut} contribution, to indicate that it is given by the cut of higher loop amplitudes.  In this expression, $r_X$ denotes the collection of  momenta $\{r_1,\cdots,r_X\}$ carried by additional particles propagating thorough the cut, whose momentum phase space integration has been explicitly indicated by $d\Phi(r_m)=\hat{d}^4r_m\hat{\delta}^{(+)}(r_m^2)$. For $n=0\,\,\rm{and}\,\,1$,  no additional photons propagate through the cut, since they only appear   starting from  $\rm{N}^2$LO in the perturbative expansion (i.e.  two-loops).

\section{Radiation Kernel : A Soft Expansion}
In this section we will study the $\mathcal{R}$  and $\mathcal{C}$-contributions to the radiated field in  the KMOC formalism at leading and subleading order in the soft expansion in 4-dimensions. We will show that consistence of the KMOC formalism with the soft theorems at the orders considered, generates a hierarchy of  constraints on the expectation value of several operators.

\subsection{Leading soft constraints }\label{sec:weinberg_kernel}

The aim of this part of the note is to  derive the set of identities  \eqref{hier1} and \eqref{hier2}. Our idea now is to use KMOC formalism in conjunction with quantum soft theorems to obtain radiation kernel in the soft limit. In other words, we start with the exact formula for the radiation kernel. We then follow the theme of  \cite{Bautista:2019tdr,Manu:2020zxl} where it was shown that taking the soft limit before the classical limit generates soft expansion of radiation kernel in KMOC formalism. That is, to a given order in soft expansion, we can apply quantum soft photon theorems to factorise the 5-point amplitude in terms of a 4 point amplitude and a soft factor. 

At higher orders in the loop expansion, one also has to take into account the order between loop integration and soft expansion.  If we first do a soft expansion and then loop integration, then one can use the tree-level soft theorems to factorise the loop integrand into a soft factor and a Four point integrand. However, as it was shown in a seminal paper by Sahoo and Sen \cite{Sahoo:2018lxl}, the two operations do not commute in Four dimensions beyond the leading order in soft expansion. That is, the soft expansion done after integrating over loop momenta results in $\ln$ soft factors  which are absent in the soft expansion of the loop integrand. 
At leading order however, this subtlety  does not enter as Weinberg soft photon theorem is a universal statement in all dimensions.

Let us then substitute the  soft photon theorem in eqns. (\ref{eq:r-kernel_}, \ref{c-k}), and use the momentum conserving delta functions to do the integrals in $q_2$ and $w_2$, we   get \footnote{We only consider the radiation emitted by the first particle, as the radiative field emitted by the outgoing particles is additive. We will denote this contribution as ${\cal R}^{\mu}_{1}(k)$.}  
\begin{equation}\label{eq:r-kernel}
  \mathcal{R}_{1}^{(n)\,\mu}(k)=i\lim_{\hbar\to0}\hbar^{\frac{3}{2}}\int\hat{d}\mu_{q}e^{-ib{\cdot}q}S^{(0)\,\mu}(p_{1},q,k)\mathcal{A}^{(n)}\left(p_{1},p_{2}{\rightarrow}p_{1}-q,p_{2}+q\right)\,.
\end{equation}
We have additionally  defined the impact parameter by  $b=b_2-b_1$, and used \eqref{eq:measure} to rewrite the momentum measure. Analogously, for the $\mathcal{C}$-term we have 
\begin{equation}\label{eq:c-kernel}
\begin{split}\mathcal{C}_{1}^{(n)\,\mu}(k) & ={\color{black}}\lim_{\hbar\to0}\hbar^{\frac{3}{2}}\sum_{X=0}^{n-1}\int\prod_{m=0}^{X}d\Phi(r_{m})\hat{d}\mu_{q}\,\hat{d}\mu_{w,X}e^{-ib{\cdot}q}\\
 & \hspace{2cm}\,\,\,\,\,\,\,\,\,\,\,\,\times\sum_{a_{1}=0}^{n-1-X}S^{(0)\,\mu}(p_{1},w,k)\mathcal{A}^{(a_{1})}(p_{1},p_{2}{\rightarrow}p_{1}{-}w,p_{2}{+}(w+r_{X}),r_{X})\\
 & \hspace{2cm}\,\,\,\,\,\,\,\,\,\,\,\,\,\times\mathcal{A}^{(n-a_{1}-X-1)\star}(p_{1}{-}w,p_{2}{+}(w+r_{X}),r_{X}{\rightarrow}p_{1}-q_{1},p_{2}-q_{2})\,,
\end{split}
\end{equation}
with $\hat{d}\mu_{w,X}$ given in \eqref{eq:measurex}.

The Weinberg soft factor has the following ``quantum" expansion when expressed in terms of exchange momenta.
For the first particle (with charge and mass $Q_{1}, m_{1}$)
\begin{flalign}\label{hexpsoft}
S^{(0) \mu}(p_{1}, p_{1}{-}q, k)\, =\, Q_{1}\, \Big[\, \sum_{i=0}^{\infty}\, q^{\alpha_{1}}\, \cdots\, q^{\alpha_{i+1}}\, V_{\alpha_{1}\, \cdots\, \alpha_{i+1}}^{\mu}\, \Big]\,,
\end{flalign}
where $V^{\mu}_{\alpha_{1}\, \cdots\, \alpha_{i+1}}$ is defined in \eqref{eq:soft_projector_1}. 
We see that the i-th term inside the square bracket in \eqref{hexpsoft}, scales as $\hbar^{i}$ in the KMOC sense.

We will now derive the constraints proposed in eqns. (\ref{hier1}) and \eqref{hier2} by associating soft limit of radiation kernel with the classical soft factor at all orders in the coupling.  The first contribution $ \mathcal{R}_{1}^{(n)\,\mu}(k)$ can be written as 
\begin{equation}\label{r-s-exp1}
\begin{split}
\mathcal{R}_{1}^{(n)\,\mu}(k)\, & =i\lim_{\hbar\to0}\hbar^{\frac{3}{2}}\,g\,Q_{1}\,g^{2(n+1)}\\
 & \,\,\,V_{1\,\alpha_{1}\,\cdots\,\alpha_{i+1}}^{\mu}\,\sum_{i=0}^{n  }\int\hat{d}\mu_{q}\,e^{-ib{\cdot}q}\,q^{\alpha_{1}}\,\cdots\,q^{\alpha_{i+1}}\,\mathcal{\bar{A}}^{(n)}\left(p_{1},p_{2}{\rightarrow}p_{1}-q,p_{2}+q\right)\,,
\end{split}
\end{equation}

where the bar over the amplitude indicates that we have striped the coupling constant. Notice here we  have restricted the sum over $i$ at $n $ (where $n$ is the order of the loop expansion). This can be argued using $\hbar$ scaling arguments. The KMOC scaling implies that
\begin{flalign}
\begin{array}{lll}
\hbar^{\frac{3}{2}}\, g^{2(n+1)}\, g\, \sim\, \frac{1}{\hbar^{n}}\,,\\
V^{\mu}_{1\,\alpha_{1}\, \cdots\, \alpha_{i+1}}\, q^{\alpha_{1}}\, \cdots\, q^{\alpha_{i+1}}\, \sim\, \hbar^{i}\,.
\end{array}
\end{flalign}

Additionally, we now notice that at $n$-th order in the loop expansion, the $\hbar$ scaling of the perturbative amplitude is quantified by KMOC as follows 
\begin{equation}
    \begin{split}
\mathcal{\bar{A}}^{(n)}\left(p_{1},p_{2}{\rightarrow}p_{1}-q,p_{2}+q\right)\, & =:\,{\cal I}^{(n)}_{4}(p_{1},p_{2}\,\rightarrow\,p_{1}-q,p_{2}+q)\\
 & \sim\,[\,\frac{1}{\hbar^{2}}\,+\,\frac{1}{\hbar}\,+\,O(\hbar^{0})\,]\,,\\
\hat{d}\mu_{q}\, & \sim\,\hbar^{4}[\frac{1}{\hbar^{2}}\,+O(\frac{1}{\hbar^{3}})\,]^{2}\,,\\
b{\cdot}q & \sim \hbar^0\,.
    \end{split}
\end{equation}

It can  be immediately verified that if  the sum $\sum_{i}$ in eqn. (\ref{r-s-exp1}) goes beyond $i = n$, the right hand side ($\hbar\, \rightarrow\, 0$ limit) vanishes. 
\textcolor{black}{In fact, these scaling arguments can be used to immediately verify that the $\hbar$ expansion of the \textit{moments} can be written as ${\cal S}_{i\, \beta}^{\mu}$, which is defined through the following equation.}
\begin{equation}\label{eq:scaling_moments_R}
g^{2(n+1)}\, g\,Q_{1}\, \hbar^{\frac{3}{2}}\, V^{\mu}_{1\, \alpha_{1}\, \cdots\, \alpha_{i+1}}\, \int\, \hat{d}^{4}\mu_q\, e^{- i b \cdot q}\, (\, q^{\alpha_{1}}\, \cdots\, q^{\alpha_{i+1}}\, )\, \, \bar{{\cal A}}^{(n)} =\, \sum_{\beta = 0}^{n -i}\, \frac{1}{\hbar^{\beta}}\, {\cal S}_{i\, \beta}^{\mu}\, + O(\hbar) \,.
\end{equation}
The contribution of $\mathcal{C}^{(n)\,\mu}(k)$ at leading order in  the soft limit can be analysed as in eqn.(\ref{r-s-exp1}). 
\begin{equation}\label{c-ker1}
\begin{split}\mathcal{C}_{1}^{(n)\,\mu}(k) & ={\color{black}}\lim_{\hbar\to0}\hbar^{\frac{3}{2}}\sum_{i=1}^{n}\, V^{\mu}_{\alpha_{1}\, \cdots\, \alpha_{i+1}}\, \sum_{X=0}^{n-1}\int\prod_{m=0}^{X}d\Phi(r_{m})\hat{d}\mu_{q}\,\hat{d}\mu_{w,X}e^{-ib{\cdot}q}\\
 & \hspace{2cm}\,\,\,\,\,\,\,\,\,\,\,\,\times\sum_{a_{1}=0}^{n-1-X}\, (w^{\alpha_{i_{1}}}\, \cdots\, w^{\alpha_{i+1}})\, \mathcal{A}^{(a_{1})}(p_{1},p_{2}{\rightarrow}p_{1}{-}w,p_{2}{+}(w+r_{X}),r_{X})\\
 & \hspace{2cm}\,\,\,\,\,\,\,\,\,\,\,\,\,\times\mathcal{A}^{(n-a_{1}-X-1)\star}(p_{1}{-}w,p_{2}{+}(w+r_{X}),r_{X}{\rightarrow}p_{1}-q_{1},p_{2}-q_{2})\,.
 \end{split}
 \end{equation}
 Once again, the scaling arguments can be used to immediately verify that the  \textit{cut} contribution to the moments at $g^{2n+3}$ order in the perturbative expansion can be written as, 
 \begin{flalign}
\begin{split}
g^{2(n+1)}\, g\,Q_{1}\, \hbar^{\frac{3}{2}}\, &V^{\mu}_{1\, \alpha_{1}\, \cdots\, \alpha_{i+1}}\, \sum_{X=0}^{n-1}\int\prod_{m=0}^{X}d\Phi(r_{m})\hat{d}\mu_{q}\,\hat{d}\mu_{w,X}e^{-ib{\cdot}q}\\
 & \hspace{2cm}\,\,\,\,\,\,\,\,\,\,\,\,\times\sum_{a_{1}=0}^{n-1-X}\, (w^{\alpha_{i_{1}}}\, \cdots\, w^{\alpha_{i+1}})\, \bar{\mathcal{A}}^{(a_{1})}(p_{1},p_{2}{\rightarrow}p_{1}{-}w,p_{2}{+}(w+r_{X}),r_{X})\\
 & \hspace{2cm}\,\,\,\,\,\,\,\,\,\,\,\,\,\times\bar{\mathcal{A}}^{(n-a_{1}-X-1)\star}(p_{1}{-}w,p_{2}{+}(w+r_{X}),r_{X}{\rightarrow}p_{1}-q_{1},p_{2}-q_{2})\\
&=\, \sum_{\beta=0}^{n -i}\, \frac{1}{\hbar^{\beta}}\, {\cal S}^{\prime\, \mu}_{i\, \beta}
 \end{split}
 \end{flalign}
 Here the sum over $X$ is constrained by the order (in the coupling) at which we are evaluating the \text{cut} contribution.
 
We finally see that for each $i$, 
 \begin{flalign}
 V^{\mu}_{\alpha_{1}\, \cdots\, \alpha_{i+1}}\, {\cal T}^{(n)\,\alpha_{1}\, \cdots\, \alpha_{i+1}}\, =\, \sum_{\beta=0}^{n -i}\, \frac{1}{\hbar^{\beta}}\, ({\cal S}_{i\, \beta} + {\cal S}_{i\, \beta}^{\prime})^{\mu}\, +\, O(\hbar)\,.
 \end{flalign}
 
 Thus, at a given order in the perturbative expansion $V^{\mu}_{\alpha_{1}\, \cdots\, \alpha_{i+1}}\, {\cal T}^{(n)\,\alpha_{1}\, \cdots\, \alpha_{i+1}}$ has a hierarchy of  super-classical terms which scales as $\frac{1}{\hbar^{\beta}}\, \vert\, \beta \in\, \{1,\, \cdots\,, n  -i\}$.  As the classical limit in KMOC formalism must be smooth, we thus conclude that to $n$-th order in the loop expansion and for each $i$, one has a tower of constraints which state that all the super-classical terms must vanish
\begin{flalign}
{\cal S}^{\mu}_{i\, \beta} + {\cal S}^{\prime}_{i\, \beta}\, =\, 0\, \forall\, \beta\, \in\, \{1,\, \cdots,\, n -i\}\,,\,\,\,\,\textcolor{black}{n>0 \, |\, 1\le i+1 \le n}.
\end{flalign}
This is precisely the first identity  \eqref{hier1}, written in a slightly different notation. 

We now analyse the classical $\beta = 0$ contribution explicitly. We can schematically write it in a form which makes the $\hbar$ scaling of various terms manifest. This can be done by isolating all the terms which do not have an $\hbar$ expansion. In particular:
(1) we separate the measure factor $\hat{d}\mu_{q}\, =\, \hat{d}^{4}q\, \hat{\delta}(q)$, and 
(2) we isolate all the measure factors over loop momenta and the $(n+1)$ massless propagators.
As can be checked, this implies that in the classical term, $\hat{\delta}(q)\, \mathcal{I}^{(n)}_4$ should scale as $\frac{1}{\hbar^{n+i-1}}$.

Let us illustrate this with ${\cal S}^{\mu}_{i\, \beta=0}\, =:\, {\cal S}^{\mu}_{i}$. 
\begin{flalign}
\begin{array}{lll}
{\cal S}^{\mu}_{i}\, =\\
g^{2(n+1)}\,g\, Q_{1}\, \hbar^{\frac{3}{2}}\, V^{\mu}_{1\, \alpha_{1}\, \cdots\, \alpha_{i+1}}\, \int\, \hat{d}^{4}q\, e^{- i b \cdot q}\, (\, q^{\alpha_{1}}\, \cdots\, q^{\alpha_{i+1}}\, )&\\  &\hspace*{-2.6in}\int\, \prod_{j=1}^{n}\hat{d}^{4}l_{j}\, \frac{1}{\prod_{m=1}^{n}l_{m}^{2}\, (\sum l_{m} - q)^{2}}\,  [\, \hat{\delta}(q)\,  {\cal I}^{(n)}_{4}(p_{1},\, p_{2}\, \rightarrow\, p_{1} - q,\, p_{2} + q)\, ]_{\frac{1}{\hbar^{n+2+i-1}}}\,.
\end{array}
\end{flalign}
One can write such a formal expression for ${\cal S}^{\mu \prime}_{i}$ analogously. 

The classical soft theorem is then a statement that $\forall\, n$ and $\forall\, 1\, \leq\, i\, \leq\, n + 1$, 
\begin{flalign}\label{csiid}
    {\cal S}^{(n)\,\mu}_{ i} + {\cal S}^{\prime (n)\, \mu}_{ i}\, =\, g^{2(n+1)}\, V^{\mu}_{s\, \alpha_{1}\, {\cdots}\, \alpha_{i+1}}\hspace{0.4in} \smashoperator{\sum_{\substack{L_{1}+\,\cdots+\,L_{i+1}=0\\
(L_{1}+L_{i+1})+(i+1)\,=n
}
}^{n+1-i}}\, (\, (\triangle p^{(L_{1})})^{\alpha_{1}}\, {\cdots}\, (\triangle p^{(L_{i+1})})^{\alpha_{i+1}}\, )\,,
\end{flalign}
which in turn recovers identity \eqref{hier2}.

In \cref{sec:tree-level-radiation} and \cref{sec:1-loop_radiation}, we verify identities \eqref{hier1} and \eqref{hier2}  up to subleading order in the perturbative expansion, i.e. $n=0$ and $n=1$. 

\subsubsection{Monomials of linear impulses}
In the previous section we expressed the soft radiation kernel as sum over certain classical \textit{moments}. Classical soft theorem implies that (expectation value) of each such \textit{moments} is sum over products of linear impulses. We can thus ask if ${\cal S}^{\mu}_{i} + {\cal S}^{\prime \mu}_{i}$ is an expectation value of certain observable. It is easy to see that the answer is indeed  affirmative. 
The tensor $V^{\mu}_{1\, \alpha_{1}\, \cdots\, \alpha_{i+1}}$ can be thought of as a map from symmetric rank $i+1$ tensor to a vector. It has a kernel spanned by $p_{1}^{\alpha_{1}}\, \cdots\, p_{1}^{\alpha_{i+1}}$. We can hence consider following quantum operators. 
Let 
\begin{flalign}
\Pi^{\mu}_{\alpha}\, =\, \delta^{\mu}_{\alpha} + \frac{1}{m_{1}^{2}}\, p_{1}^{\mu}\, p_{1 \alpha}\,.
\end{flalign}
Now consider a quantum operator, 
\begin{flalign}
{\cal P}_{1}^{\mu}\, =\, \Pi^{\mu}_{1\, \alpha}\, \hat{P}_{1}^{\alpha}\,,
\end{flalign}
with $\hat{P}_1$ the momentum operator for particle 1. 
The identities (given in eqn.(\ref{csiid}) implied by consistency with classical soft theorem is then a statement that 
\begin{flalign}
\begin{array}{lll}
\lim_{\hbar\rightarrow\, 0}\, V^{\mu}_{1\, \alpha_{1}\, \cdots\, \alpha_{i+1}}\langle\langle\, {\cal P}^{\alpha_{1}}\, \cdots\, {\cal P}^{\alpha_{i+1}}\, \rangle\rangle^{(n)}\ =
V^{\mu}_{s\, \alpha_{1}\, {\cdots}\, \alpha_{i+1}}\, \smashoperator{\sum_{\substack{L_{1},\,\cdots,\,L_{i+1}=0\\
(L_{1}+L_{i+1})+(i+1)\,=n
}
}^{n+1-i}}\, (\, (\triangle p^{(L_{1})})^{\alpha_{1}}\, {\cdots}\, (\triangle p^{(L_{i+1})})^{\alpha_{i+1}}\, )
\end{array}\,.
\end{flalign}

\subsection{Towards  Sub-leading soft constraints}\label{sec:subleading_soft_sonstraints}
In this section we sketch the possible constraints that arise at the sub-leading order in the soft expansion due to the classical log soft photon theorem. That is, we use the quantum sub-leading soft theorem to evaluate the 5 (or higher) point amplitude in the radiation kernel and then take the classical limit. We then equate the sub-leading (terms which scale as $\ln\omega$) soft radiation kernel with classical log soft factor order by order in perturbation theory and as we argue, this generates another hierarchy of constraints on certain moments involving exchange momenta and angular momentum operators. 

A detailed proof of these constraints is outside the scope of this work and will be pursued elsewhere. For now, we simply sketch the structural forms of these constraints. 

In four dimensions, the soft expansion and loop integration do not commute beyond leading order due to infra-red divergences \cite{Sahoo:2018lxl}. Although in the classical limit, the loop integral is expected to only produce IR-finite quantities, we use the soft expansion of the loop integrand to compute the radiation kernel. At the level of loop integrand, the soft expansion is simply the tree-level expansion and hence we can use the sub-leading soft photon factor to proceed. 

More in detail, we write the $5$ point amplitude under consideration as,
\begin{flalign}
\begin{array}{lll}
{\cal A}_{5}^{(n)\, \mu}(p_{1}, p_{2}\, {\rightarrow}\, p_{1} {-} q_{1}, p_{2} {-} q_{2},k)\, \delta^{4}(q_{1} {+} q_{2}{-}k)\, =\\
\hspace{2cm}
\int_{l_{1},{\cdots}, l_{n}}\, \sum_{i=1}^{2} Q_{i} \Big[\, \frac{J_{i}^{\mu\nu}k_{\nu}}{p_{i} \cdot k}\, {+}\, \frac{\tilde{J}_{i}^{\mu\nu}k_{\nu}}{\tilde{p}_{i} \cdot k}\, \Big] \delta^{4}(q_{1} + q_{2})\,I_{4}(p_{1}, p_{2}, q_{1}, q_{2}, l_{1}, {\cdots}, l_{n})\,,
\end{array}
\end{flalign}
where $J_i^{\mu\nu} = (p_i\wedge\partial_{p_i})^{\mu\nu}$, is the angular momentum operator for the incoming particles, and with  similar expression for the  outgoing ones $\tilde{p}_i = p_i-q_i$. In addition,  $I_{4}$ is the loop integrand. Here we have to note that $J_i^{\mu\nu} $ acts of both, the momentum conserving delta function, as well as the striped integrand. 

Using this expansion, the $\mathcal{R}$-part of the radiated field \eqref{eq:r-kernel}  at subleading order in the soft expansion becomes 
\begin{equation}
\begin{array}{lll}
    \mathcal{R}^{(n)\,\mu}(k)=
    i\lim_{\hbar\to0}\hbar^{\frac{3}{2}}\int\prod_{i=1}^{2}\hat{d}^{4}q_{i}\hat{\delta}(2p_{i}{\cdot}q_{i}-q_{i}^{2})e^{ib_{i}{\cdot}q_{i}}\\
  \hspace{2cm}  \times\int_{l_{1}, {\cdots}, l_{n}}\, \sum_{i=1}^{2} Q_{i} \Big[ \frac{J_{i}^{\mu\nu}k_{\nu}}{p_{i} \cdot k}+ \frac{\tilde{J}_{i}^{\mu\nu}k_{\nu}}{\tilde{p}_{i} \cdot k}\Big] \delta^{4}(q_{1} {+} q_{2})\,I_{4}(p_{1}, p_{2}, q_{1}, q_{2}, l_{1},{ \cdots}, l_{n})\,.
    \end{array}
\end{equation}
As before when we expand $\frac{1}{(p_{i} + q_{i})\cdot k}$ in $\hbar$ expansion by expanding in powers of $\frac{q_{i} \cdot k}{p_{i} \cdot k}$, it can be verified that the only terms that will survive in the classical limit are,
\begin{equation}\label{sscq}
\begin{array}{lll}
    \mathcal{R}^{(n)\,\mu}(k)&=
    i\lim_{\hbar\to0}\hbar^{\frac{3}{2}}\int\prod_{i=1}^{2}\hat{d}^{4}q_{i}\hat{\delta}(2p_{i}{\cdot}q_{i}-q_{i}^{2})e^{ib_{i}{\cdot}q_{i}}\\&
    \hspace{0.5cm}\times
    \int_{l_{1},{\cdots}, l_{n}}\, \sum_{i=1}^{2} Q_{i}\,  
    \Big[\, \frac{(\tilde{J}_{i}{+} J_{i})^{\mu\nu}k_{\nu}}{p_{i} \cdot k} {+} \sum_{m=1}^{n}\, ({-}1)^{m}(\frac{q_{m} {\cdot} k}{p_{m}{\cdot}k})^{m}\, \tilde{J}_{i}^{\mu\nu}k_{\nu} \Big]\\&\hspace{3cm}\times
    \delta^{4}(q_{1} {+} q_{2})\,I_{4}(p_{1}, p_{2}, q_{1}, q_{2}, l_{1},{\cdots}, l_{n})\,.
    \end{array}
\end{equation}
One can similarly apply the (tree-level) sub-leading soft photon theorem in ${\cal C}^{(n) \mu}(k)$.

Consistency with classical soft theorem then implies that in  the sub-leading soft radiation kernel:  (1) all the super-classical terms vanish, and (2) 
there must exist integration regions in all of which the exchange momenta lie in $b^{-1}\, >>\, \vert q\, \vert >>\, \omega$  that generate various terms in the radiative field proportional to $\ln\omega$. A detailed analysis of the loop integration region  is under investigation and will be pursued elsewhere.\footnote{e.g. We suspect that at n-th order in the perturbation theory,  the contribution proportional to $(\triangle J^{\ln})^{(n)}$ may arise from $\omega\, <<\, \vert l_{i}\vert\, <<\, \vert q\vert\, <<\, b^{-1}$. However this remains to be shown.} 
As we argued previously, the log-soft radiative field can be perturbatively expanded as, 
\begin{flalign}\label{sscc}
\begin{array}{lll}
A_{\mu}^{\ln (n)}\, =
\sum_{m}\, Q_{m}\, \Big[\, \frac{1}{p_{m} \cdot k}(\triangle J^{\ln}_{m})^{(n) }_{\mu\nu}\, k^{\nu}\,+\\ \hspace{1cm} \sum_{i=1}^{n}\, \, \smashoperator{\sum_{\substack{L_{1},{\cdots}, L_{i}\\ L_{1} {+} {\cdots} L_{i} {+} L_{i+1} {+} i{+}1 {=} n}}} \,\,\,\,\frac{(-1)^{i}}{(p_{m} {\cdot} k)^{i+1}}\, k_{\alpha_{1}}\, k_{\alpha_{i}}\, (\, (\triangle p^{L_{1}}_{m})^{\alpha_{1}}\, \cdots=0\, (\triangle p^{L_{i}}_{m})^{\alpha_{i}}\, )\, (\triangle J^{\ln\, (L_{i+1})}_{m})_{\mu\nu}\, k^{\nu}\, \Big]\,,
\end{array}
\end{flalign}
where $(\triangle J^{\ln})^{(i)}$ is the logarithmic divergent term in angular momentum impulse at N$^{i}$LO order in the coupling. 

Equality between the right hand side of eqn.(\ref{sscc}) and the $\ln\omega$ contribution from the right hand side of eqn.(\ref{sscq}) (plus the contribution from the $\mathcal{C}$-term),  provides the  sub-leading soft constraints analogs of (\ref{hier1}-\ref{hier2}).
At leading order in the coupling (that is $n=0$), these constraints were verified in \cite{Manu:2020zxl}.  At the NLO order, we expect the sub-leading soft kernel from KMOC side to match the results in  \cref{sec:subleading_soft_factor_classical}. \cite{Baut-Lad}. 

\section{Leading Soft Constraints Verification up to NLO }\label{sec:lo-soft_costraitns}
Let us in the remaining of this note to provide some specific tests for identities \eqref{hier1} and \eqref{hier2}, at leading ($n=0$) and subleading ($n=1$) orders in perturbation theory. 

\subsection{Tree-level leading soft \textit{moments}}\label{sec:tree-level-radiation}

At tree-level  there is not superclassical term and therefore \eqref{hier1} does not impose any constrain. On the other hand, the classical   \textit{moment} contributing at this order is  ${\cal T}^{(0)\,\alpha}$ in \eqref{eq:moments}. This in turn will allow us to recover the radiated field \eqref{eq:leading-soft-tree} at leading order in perturbation theory. In other words, in the classical limit  we just need to show that $\lim_{\hbar\to0}{\cal T}^{(0)\,\alpha}= g^2\,\Delta p_1^{(0),\alpha}$ as required by \eqref{hier2}. We remark  that this, and the equalities written below, are only valid on the support of the projector $V^\mu_{\alpha_1,\cdots,\alpha_i+1}$, as indicated in (\ref{hier1}-\ref{hier2}).

For leading order radiation, as already mentioned, only the first term in \eqref{eq:moments} contributes to the computation of the \textit{moment}, since the remaining term is subleading in the coupling.  Additionally, since we are taking the classical limit, the following expansion for the momentum measure \eqref{eq:measure} will be useful for us. 
\begin{align}
\hat{d}\mu_q & =\hat{d}\mu_{1\,q}+\hat{d}\mu_{2\,q}+\cdots,\label{eq:full measure}\\
\hat{d}\mu_{1\, q} & =\hat{d}^{4}q\,\hat{\delta}(2p_{1}{\cdot}q)\hat{\delta}(2p_{2}{\cdot}q),\label{eq:hbar0 measure}\\
\hat{d}\mu_{2\,q }& =-\,\hat{d}^{4}q\,q^{2}\left[\hat{\delta}'(2p_{1}{\cdot}q)\hat{\delta}(2p_{2}{\cdot}q)-\hat{\delta}(2p_{1}{\cdot}q)\hat{\delta}'(2p_{2}{\cdot}q)\right].\label{eq:hbar1measure}
\end{align}
We will also need the classical piece of the  tree-level 4 point amplitude, which is   given by the one photon exchange diagram
\begin{equation}\label{eq:4pt_LO}
    \mathcal{A}^{(0)}(p_1,p_2\to p_1-q,p_1+q)  = 4 g^2 Q_{1}Q_{2}\frac{\,p_{1}{\cdot}p_{2}}{q^{2}+i\epsilon}\,.
\end{equation}

With all these ingredients at hand, the only non-vanishing contribution to the   \textit{moment} ${\cal T}^{(0)\,\alpha}$ in the classical limit,  can be obtained by replacing  \eqref{eq:hbar0 measure} and \eqref{eq:4pt_LO} into \eqref{two}, after which it follows 

\begin{equation}\label{eq:moment0}
\begin{split}
  \lim_{\hbar\, \rightarrow\, 0}\, V^{\mu}_{\alpha}  
    {\cal T}^{(0)\,\alpha} &=  g^2V^{\mu}_{\alpha}   \int \hat{d}^4q \hat{\delta}(p_1{\cdot}q)\hat{\delta}(p_2{\cdot}q) \frac{iQ_1Q_2\,p_1{\cdot}p_2q^\alpha}{q^2+i\epsilon} e^{-i q{\cdot}b}\,,\\
    & =g^2V^{\mu}_{\alpha}   \Delta p_1^{(0)\,\alpha}\,,
   \end{split} 
\end{equation}
which indeed satisfies the identity \eqref{hier2} for $n=0$. In the second line we have identified  the integral representation for the leading order  impulse 
\begin{equation}\label{eq:LO_impulse_integral}
    \Delta p^{(0)\,\mu}_{1} =  \int \hat{d}^4q \hat{\delta}(p_1{\cdot}q)\hat{\delta}(p_2{\cdot}q) \frac{iQ_1Q_2\,p_1{\cdot}p_2q^\mu}{q^2+i\epsilon} e^{-i q{\cdot}b}.
\end{equation}

To finish this example, let us explicitly evaluate this integral, even though it has been evaluated in several previous works (see for instance \cite{Kosower:2018adc,Guevara:2018wpp}). We aim to introduce  some conventions that will be used in the remaining of the paper. 

We start by noticing that since there are two delta functions that allow us to evaluate the integrals in the time and longitudinal directions,  the $i\epsilon$  prescription for the propagator  is irrelevant, which in turn implies that  the result for the impulse will be the same irrespective of whether we used the Feynman or the Retarded propagator \footnote{However, this will not be the case for all of the integrals that we will find in this paper as we will see below. }. Moving forward in the computation, we can now  decompose the momentum $q$ in terms of  the massive momenta $p_i$, and the transverse momentum $q_\perp$, as follows
\begin{equation}\label{eq:momentum-expansion-q}
    q^\mu=\alpha_{2}p_{1}^\mu+\alpha_{1}p_{2}^\mu+q_{\perp}^\mu,\quad p_{i}{\cdot}q_{\perp}=0,
\end{equation}
where 
\begin{equation}\label{eq:alphai_factors}
    \alpha_{1}=\frac{1}{\mathcal{D}}\left[p_{1}{\cdot}p_{2}x_1{-}m_{1}^{2}x_2\right],\quad \alpha_{2}=\frac{1}{\mathcal{D}}\left[p_{1}{\cdot}p_{2}x_2{-}m_{2}^{2}x_1\right].
\end{equation}
Here we have introduced the dimension-full quantities $x_i  = p_i{\cdot}q$, and the Jacobian factor  $\mathcal{D}$, given by \eqref{eq:jacobian}.

Notice that the  decomposition \eqref{eq:momentum-expansion-q} is generic and does not assume any conditions on the $x_i$ variables.
With this change of variables, the integral measure in \eqref{eq:LO_impulse_integral} becomes $ \hat{d}^{4}q=\frac{1}{\sqrt{\mathcal{D}}}\hat{d}^{2}q_{\perp}\hat{d}x_1\hat{d}x_2$.

In general we will have to evaluate integrals of the form 
\begin{equation}\label{eq:general_integral}
    \mathcal{I} = \frac{1}{\sqrt{\mathcal{D}}}\int \hat{d}^{2}q_{\perp}\hat{d}x_1\hat{d}x_2 \hat{\delta}^{(n)}(x_1)\hat{\delta}^{(m)}(x_2) f(x_1,x_2,q_\perp,\sigma),
\end{equation}
namely, with  a certain number of derivatives acting over  the on-shell delta functions. We can  use integration by part multiple times in order to remove the derivatives acting over the delta functions,  transporting them to act over the integrand function $f(x_1,x_2,q_\perp,\sigma)$ \footnote{ Here  we have use $\sigma$ to represent additional momenta, masses and impact parameter labels.}; once we have the on-shell delta functions free of derivatives, we can use them to evaluate the $x_i$-integrals. At that point, the calculation  would have been  reduced  to  evaluate the lower-dimensional integrals  of the form
\begin{equation}\label{eq:general_ibp}
    \mathcal{I} = (-1)^{m+n}\frac{1}{\sqrt{\mathcal{D}}}\int \hat{d}^{2}q_{\perp} \frac{\partial^n}{\partial x_1^n}\frac{\partial^m}{\partial x_2^m}
    f(x_1,x_2,q_\perp ,\sigma)\Bigg|_{x_1=x_2=0}.
\end{equation}

Let us go back to the computation of  the leading order impulse integral \eqref{eq:LO_impulse_integral}. For this case,  $m=n=0$, and therefore  the evaluation of the  integrals in the time and longitudinal directions simply reduces to  set $\alpha_1=\alpha_2=0$.  We are left then  with the two-dimensional integral 
\begin{equation}\label{eq:LO_impulse_interm}
\Delta p^{(0)\,\mu}_{1} =\frac{Q_{1}Q_{2}\,p_{1}{\cdot}p_{2}}{\sqrt{\mathcal{D}}}\int \hat{d}^{2} q_\perp e^{-iq_\perp{\cdot}b} \frac{iq_\perp^{\mu}}{q_\perp^{2}}\,,
\end{equation}
which can be evaluated by trading  the momentum $q_\perp$ in the numerator by a derivative w.r.t the impact parameter. Afterwards, the integral can be evaluated in polar coordinates
\begin{align}
\Delta p^{(0)\,\mu}_{1}  & =\frac{Q_{1}Q_{2}\,p_{1}{\cdot}p_{2}}{\sqrt{\mathcal{D}}}\frac{1}{2\pi}\partial_{b^{\mu}}\lim_{\mu\to0}\int_{\mu}^{\infty}\frac{dq_{\perp}}{q_{\perp}}\int_{0}^{2\pi}\frac{d\theta}{2\pi}e^{iq_{\perp}b_{\perp}\cos\theta}\,,\\
 & =\frac{Q_{1}Q_{2}\,p_{1}{\cdot}p_{2}}{\sqrt{\mathcal{D}}}\frac{1}{2\pi}\partial_{b^{\mu}}\lim_{\mu\to0}\int_{\mu}^{\infty}dq_{\perp}\frac{\mathcal{J}_{0}(q_{\perp}b_{\perp})}{q_{\perp}}\,,\\
 & =-\frac{1}{4\pi}\frac{Q_{1}Q_{2}\,p_{1}{\cdot}p_{2}}{\sqrt{\mathcal{D}}}\lim_{\mu\to0}\partial_{b^{\mu}}\ln\left(-b^{2}\mu^{2}\right)\label{eq:IR_impulse_interm}\,.
\end{align}
In the second line $\mathcal{J}_0(x)$ is the order zero Bessel functions of the first kind.
Taking the derivative and trivially evaluating the $\mu \to 0$ limit,  leads to the well know result for the leading order electromagnetic impulse
\begin{equation}\label{eq:LO_Impulse_explicit}
   \Delta p^{(0)\,\mu}_{1}=-\frac{Q_{1}Q_{2}\,p_{1}{\cdot}p_{2}}{2\pi\sqrt{\mathcal{D}}}\frac{b^{\mu}}{b^{2}},\,\,\,\,\,b^{2}=-\vec{b}^{2},
\end{equation}
where $\vec{b}$ is the two dimensional impact parameter. This concludes the computation for the radiated photon field at leading order in both, the soft, and the perturbative expansions. Let us now do the analogous computation at NLO in perturbation theory. 

\subsection{One-loop leading soft \textit{moments}}\label{sec:1-loop_radiation}

At NLO in the perturbative  expansion  the contributing 
\textit{moments} are  ${\cal T}^{(1)\,\alpha}$ and  ${\cal T}^{(1)\,\alpha\,\beta}$. In this section we want to show  that $\lim_{\hbar\to0}{\cal T}^{(1)\,\alpha} = g^4 \Delta p^{(1)\,\alpha}_{1} $, recovering  the NLO impulse, whereas $ \lim_{\hbar\to0}{\cal T}^{(1)\,\alpha\,\beta} = g^4 \Delta p^{(0)\,\alpha}_{1} \Delta p^{(0)\,\beta}_{1}$, as suggested by the second identity  \eqref{hier2}. Combination of these two results allow us to recover the one loop contribution to the  radiated field given explicitly in 
\eqref{eq:soft-one-loop}.

At NLO,  the radiated field scales as $g^5$ and therefore
the \textit{moments} receive contributions from both the $\mathcal{R}$ and the $\mathcal{C}$ terms, given by the first and second line of \eqref{eq:moments}, respectively. However, at this order no  extra photons propagate through  the cut and we can simply set $X=0$  in \eqref{eq:moments}, which also implies that $\hat{d}\mu_{w,X}=\hat{d}\mu_{w}$ in \eqref{eq:measurex}. 
In addition, we  will  show that superclassical terms give vanishing contribution as suggested by the first identity  \eqref{hier1}. Indeed, this corresponds to  a cancellation between the  $\mathcal{R}$ and the $\mathcal{C}$ contributions to the aforementioned  \textit{moments}, which is analogous to that for the computation of the  at 1-loop impulse  \cite{Kosower:2018adc}. Since only the  \textit{moment}  ${\cal T}^{(1)\,\alpha}$ will have potential superclassical contributions, coming from the superclassical piece of the 4 point amplitude at 1-loop \cite{Kosower:2018adc}, we only have to show that for $m=1$, $\lim_{\hbar\to0}\hbar^m\,{\cal T}^{(1)\,\alpha}=0$, as for higher values of $m$, identity 1 is trivially satisfied.

Let us split the computation as follows: For the potentially superclassical contributions we will compute

\begin{equation}\label{eq:super_class_total}
    \lim_{\hbar\to0}\hbar\,V_\alpha^\mu{\cal T}^{(1)\,\alpha} = \lim_{\hbar\to0}\,V_\alpha^\mu\Big[ {\cal T}^{(1)\,\alpha}_{\mathcal{R}_0} +   {\cal T}^{(1)\,\alpha}_{\mathcal{C}_0}\Big]\,,
\end{equation}
where
\begin{eqnarray}
{\cal T}^{(1)\,\alpha}_{\mathcal{R}_0}  & =& i \hbar^{5/2} \int\hat{d}\mu_{1\,q}\, e^{-ib{\cdot}q}\,q^\alpha\,\mathcal{A}^{(1)}_{\rm{sc}}(q)\,,\label{eq:R0-1} \\
 {\cal T}^{(1)\,\alpha}_{\mathcal{C}_0} & =&  \hbar^{5/2}
 \int\hat{d}\mu_{1\,q}\,\hat{d}\mu_{1\,w}\,e^{-ib{\cdot}q}\,w^\alpha \mathcal{A}^{(0)\,\star}(w-q)\mathcal{A}^{(0)}(w)\,.\label{eq:C0 definition-2}
\end{eqnarray}
Here $\mathcal{A}^{(1)}_{\rm{sc}}(q)$ is the superclassical piece of the 1-loop, 4 point amplitude, which we will write explicitly below. The tree level amplitudes in the second line are  given by \eqref{eq:4pt_LO}, where we have removed the massive momenta labels to alleviate notation.

Next, we will have to compute the classical  contributions, from the one and two index moment. For the former we have 
\begin{equation}\label{eq:1_index_moment_one_loop}
    \lim_{\hbar\to0}\,V_\alpha^\mu{\cal T}^{(1)\,\alpha} = \lim_{\hbar\to0}\,V_\alpha^\mu\Big[ {\cal T}^{(1)\,\alpha}_{\mathcal{R}_1}+{\cal T}^{(1)\,\alpha}_{\mathcal{R}_2} +   {\cal T}^{(1)\,\alpha}_{\mathcal{C}_1}+  {\cal T}^{(1)\,\alpha}_{\mathcal{C}_2}\Big]\,,
\end{equation}
with each term computed as follows
\begin{eqnarray}
 {\cal T}^{(1)\,\alpha}_{\mathcal{R}_1} &= & i \hbar^{3/2} \int\hat{d}\mu_{1\,q}\, e^{-ib{\cdot}q}\,q^\alpha\,\mathcal{A}^{(1)}_{\rm{c}}(q)\,,\label{eq:R1-1}\\
{\cal T}^{(1)\,\alpha}_{\mathcal{R}_2}  &=&i \hbar^{3/2}\, \int\hat{d}\mu_{2\,q}\,e^{-ib{\cdot}q} \,q^\alpha \mathcal{A}^{(1)}_{\rm{sc}}(q)\,,\label{eq:R2-1}\\
 {\cal T}^{(1)\,\alpha}_{\mathcal{C}_1} &=&   \hbar^{3/2}
 \int\hat{d}\mu_{1\,q}\,\hat{d}\mu_{2\,w}\,e^{-ib{\cdot}q}\,w^\alpha \mathcal{A}^{(0)\,\star}(w-q)\mathcal{A}^{(0)}(w)\,,\label{eq:C1 definition-2} \\
 {\cal T}^{(1)\,\alpha}_{\mathcal{C}_2} &=&  \hbar^{3/2}
 \int\hat{d}\mu_{2\,q}\,\hat{d}\mu_{1\,w}\,e^{-ib{\cdot}q}\,w^\alpha \mathcal{A}^{(0)\,\star}(w-q)\mathcal{A}^{(0)}(w)\,.\label{eq:C2 definition-2} 
\end{eqnarray}
In the first line, $\mathcal{A}^{(1)}_{\rm{c}}(q)$ is the classical part of the 1-loop 4 point amplitude, which we  will write  explicitly in a moment. 

Finally, the classical contribution from the  two-index \textit{moment} will be computed from
\begin{equation}\label{eq:two-index-moment-1-loop}
     \lim_{\hbar\to0}\,V_{\alpha\,\beta}^\mu{\cal T}^{(1)\,\alpha\,\beta} = \lim_{\hbar\to0}V_{\alpha\,\beta}^\mu\,\Big[ {\cal T}^{(1)\,\alpha\,\beta}_{\mathcal{R}_3}+   {\cal T}^{(1)\,\alpha\,\beta}_{\mathcal{C}_3} \Big]\,,
\end{equation}
with the respective terms evaluated via
\begin{eqnarray}
  {\cal T}^{(1)\,\alpha\,\beta}_{\mathcal{R}_3} &=&  i \hbar^{3/2} \int\hat{d}\mu_{1\,q}\, e^{-ib{\cdot}q}\,q^\alpha\,q^\beta \,\mathcal{A}^{(1)}_{\rm{sc}}(q)\,,\label{eq:R3-1} \\
  {\cal T}^{(1)\,\alpha\,\beta}_{\mathcal{C}_3} &=&  \hbar^{3/2}
 \int\hat{d}\mu_{1\,q}\,\hat{d}\mu_{1\,w}\,e^{-ib{\cdot}q}\,w^\alpha\,w^\beta \mathcal{A}^{(0)\,\star}(w-q)\mathcal{A}^{(0)}(w)\,,\label{eq:C3 definition-2}
\end{eqnarray}

By explicit evaluation,  we will show that the actual terms contributing to  the radiated photon field are  \eqref{eq:R1-1}, \eqref{eq:R2-1} and  \eqref{eq:R3-1} -- as suggestively written  in \eqref{eq:scaling_moments_R} --  with the first two giving the NLO impulse, and the last one giving the square of the leading order impulse. As for the   remaining contributions we show that they canceling among themselves. In what follows we will adventure  in this  computation.

\subsubsection{The superclassical terms}
Let us start by computing the superclassical terms   \eqref{eq:R0-1} and \eqref{eq:C0 definition-2}.  As we will see, these terms are IR divergent, in analogy to the IR divergent integrals appearing in the computation of the 2PM two-body potential \cite{Cheung:2018wkq,Bern:2019crd}, and the cancellation here is the KMOC analog of the cancellation for the EFT and full theory amplitudes matching \cite{Cheung:2018wkq,Bern:2019crd}. Indeed, we will see that analogous comparisons follow for the different terms appearing in the 2PM two-body potential as we will see below

The 4 point amplitude at 1-loop was computed in \cite{Kosower:2018adc}. The superclassical contribution $ \mathcal{A}^{(1)}_{\rm{sc}}(q)$, arises from the addition of superclassical parts in the box $B_{-1}$, and cut-box $C_{-1}$ diagrams, given by  eq. $(5.31)$ in \cite{Kosower:2018adc}, 
\begin{equation}\label{eq:A4_super_classical}
   \mathcal{A}^{(1)}_{\rm{sc}}(q) = \left(B_{-1}+C_{-1}\right)_{\hbar^{-1}}=2i\,g^4\left(Q_{1}Q_{2}\,p_{1}{\cdot}p_{2}\right)^{2}\int\hat{d}^{4}l\prod_{i}\hat{\delta}(p_{i}{\cdot}l)\frac{1}{l^{2}(l-q)^{2}}\,.
\end{equation}
Using  it into  \eqref{eq:R0-1}, together with the measure  \eqref{eq:hbar0 measure},  \eqref{eq:R0-1} becomes
\begin{equation}\label{eq:supcR}
{\cal T}^{(1)\,\alpha}_{\mathcal{R}_0} =  - \frac{1}{2}g^4\left(Q_{1}Q_{2}\,p_{1}{\cdot}p_{2}\right)^{2} \int\hat{d}^{4}l\hat{d}^{4}q\prod_{i}\hat{\delta}(p_{i}{\cdot}l)\hat{\delta}(p_{i}{\cdot}q)\frac{q^{\alpha}}{l^{2}(l-q)^{2}}e^{-ib{\cdot}q}\,,
\end{equation}
where we see that the explicit dependence in $\hbar$ drops away by using the KMOC $\hbar$-rescaling mentioned in \cref{sec:review_KMOC}.  We can now do the change of variables $q=l+\bar{q}$. This in turn factorizes the integrals into two factors corresponding to a vector, and a scalar integrals; that is
\begin{equation}\begin{split}
{\cal T}^{(1)\,\alpha}_{\mathcal{R}_0} =ig^4\left(Q_{1}Q_{2}\,p_{1}{\cdot}p_{2}\right)^2S_{\alpha,\omega^{-1}}^{\mu}\left[\int\hat{d}^{4}\bar{q}\prod_{i}\hat{\delta}(p_{i}{\cdot}q)e^{-ib{\cdot}\bar{q}}\frac{i\bar{q}^{\alpha}}{\bar{q}^{2}}\right]
 \left[\int\hat{d}^{4}l\prod_{i}\hat{\delta}(p_{i}{\cdot}l)\frac{e^{-ib{\cdot}q}}{l^{2}}\right],
\end{split}
\label{eq:super classical intermediate}
\end{equation}
where  the change of variables has produced a factor of $2$ that
canceled the  $\frac{1}{2}$ overall factor in \eqref{eq:supcR} \footnote{Note that formally the change of variables implies that  we should
had changed $\hat{\delta}(p_{i}.q)\rightarrow\hat{\delta}(p_{i}.\bar{q}-p_{i}{\cdot}l),$
however, the delta functions $\hat{\delta}(p_1{\cdot}q)$ allow us to set
$p_{i}{\cdot}l\rightarrow0$.}. In the integral on the left, we recognize the leading order impulse $(\ref{eq:LO_impulse_integral})$, whereas  for the integral  on
the right, we obtain an IR-divergent expression, which can be evaluated along  similar steps used for the  computation of the leading order impulse \eqref{eq:IR_impulse_interm}, obtaining

\begin{equation}
I_{1}=\int\hat{d}^{4}l\prod_{i}\hat{\delta}(p_{i}{\cdot}l)\frac{e^{-ib{\cdot}l}}{l^{2}}=-\frac{1}{4\pi\sqrt{\mathcal{D}}}\ln\left(-\mu^{2}b^{2}\right)\,.\label{eq:scalar integral I1}
\end{equation}
Here we have   introduced the   IR-regulator $\mu$. Then,  the first superclassical contribution becomes 
\begin{equation}
{\cal T}^{(1)\,\alpha}_{\mathcal{R}_0}=-ig^4 \frac{Q_{1}Q_{2}\,p_{1}{\cdot}p_{2}}{4\pi\sqrt{\mathcal{D}}} \,\Delta p_{1}^{(0)\,\alpha}\,\ln\left(-\mu^{2}b^{2}\right),\label{eq:super classical final}
\end{equation}

Let us now evaluate the  \textit{cut} contribution \eqref{eq:C0 definition-2}. For that we just need the tree-level 4 point amplitude \eqref{eq:4pt_LO}, as well as the measure factors \eqref{eq:hbar0 measure}; we arrive at
\begin{equation}
   {\cal T}^{(1)\,\alpha}_{\mathcal{C}_0}  =16g^4\left(Q_{1}Q_{2}p_{1}{\cdot}p_{2}\right)^{2}\,\int\hat{d}^{4}q\hat{d}^{4}l\prod_i\hat{\delta}(2p_{i}{\cdot}q)\hat{\delta}(2p_{i}{\cdot}l)e^{-ib{\cdot}q}\frac{l^\alpha}{l^{2}(q-l)^{2}}\,.
\end{equation}
After doing the same change of variables  $q=l+\bar{q}$, we can analogously identify the leading order  impulse from the $l$-integral, whereas the $\bar{q}$-integral will result into the  IR-divergent expression  \eqref{eq:scalar integral I1}. We finally get
\begin{equation}
     {\cal T}^{(1)\,\alpha}_{\mathcal{C}_0}  =ig^4\frac{Q_{1}Q_{2}\,p_{1}{\cdot}p_{2}}{4\pi\sqrt{\mathcal{D}}} \,\Delta p_{1}^{(0)\,\alpha}\,\ln\left(-\mu^{2}b^{2}\right),
\end{equation}
which is equal to \eqref{eq:super classical final} but with opposite sign. This explicitly  shows that the r.h.s of \eqref{eq:super_class_total} evaluates to  zero,  as demanded from the first identity  \eqref{hier1}.

\subsubsection{Classical one-index \textit{moment} at 1-loop}\label{sec:one-index-1-loop}
Let us move to evaluate the classical contribution from the one-index \textit{moment} \eqref{eq:1_index_moment_one_loop}. We start from term \eqref{eq:R1-1}. For that,  we need the classical contribution to 4 point  amplitude at 1-loop. Likewise for the superclassical term, we obtain it from the sum  $(B_{0}+C_{0})+\left(B_{-1}+C_{-1}\right)+T_{12}+T_{21}$, where the different components where evaluated in eqs.  $(5.21)$ and $\,(5.34)$  in \cite{Kosower:2018adc}. This gives

\begin{align}
\mathcal{A}_{\textrm{c}}^{(1)}(q) & =2g^4\left(Q_{1}Q_{2}\,p_{1}{\cdot}p_{2}\right)^{2}\int\frac{\hat{d}^{4}l}{l^{2}(l{-}q)^{2}}\Bigg\{ l{\cdot}(l{-}q)\left[\frac{\hat{\delta}(p_{2}{\cdot}l)}{(p_{1}{\cdot}l{+}i\epsilon)^{2}}{+}\frac{\hat{\delta}(p_{1}{\cdot}l)}{(p_{2}{\cdot}l{-}i\epsilon)^{2}}\right]\nonumber \\
 & +\frac{1}{\left(p_{1}{\cdot}p_{2}\right)^{2}}\left[m_{2}^{2}\hat{\delta}(p_{2}{\cdot}l){+}m_{1}^{2}\hat{\delta}(p_{1}{\cdot}l)\right]\Bigg\}+Z,\label{eq:4pt 1L classical explicit}
\end{align}
with
\begin{equation}
Z=ig^4\left(Q_{1}Q_{2}\,p_{1}{\cdot}p_{2}\right)^{2}\int\frac{\hat{d}^{4}l}{l^{2}(l{-}q)^{2}}(2l{\cdot}q{-}l^{2})\left[\hat{\delta}'(p_{1}{\cdot}l)\hat{\delta}(p_{2}{\cdot}q){-}\hat{\delta}(p_{1}{\cdot}l)\hat{\delta}'(p_{2}{\cdot}l)\right].\label{eq:Z mu contribution to boxes}
\end{equation}

Before we proceed with the computation, let us first  remember the definition for the electromagnetic  impulse at 1-loop \cite{Kosower:2018adc} :

\begin{equation}
\Delta p^{(1)\,\mu}_{1}=\frac{i}{4}\int\hat{d}^{4}q\prod_{i}\hat{\delta}(p_{i}{\cdot}q)e^{-ib{\cdot}q}\left[\mathcal{I}_{1}^{\mu}+\mathcal{I}_{2}^{\mu}+\mathcal{I}_{3}^{\mu}\right],\label{eq:NLO impulse}
\end{equation}
 where the $\mathcal{I}_{i}^{\mu}$ integrals resemble the contributions
to the 4 point  amplitude from the different Feynman diagrams. The firs
one comes from the contribution from the triangle diagrams 
\begin{equation}
\mathcal{I}_{1}^{\mu}=2g^4\left(Q_{1}Q_{2}\right)^{2}q^{\mu}\sum_{i}\int\hat{d}^{4}l\frac{m_{i}^{2}\hat{\delta}(p_{i}{\cdot}l)}{l^{2}(l-q)^{2}}\,.\label{eq:I1}
\end{equation}
 Next we have the contribution coming from the Boxes, which once by canceling the term 
$Z$ in $(\ref{eq:4pt 1L classical explicit})$, using the cut-box  reads 
\begin{equation}
\mathcal{I}_{2}^{\mu}=2g^4\left(Q_{1}Q_{2}p_{1}{\cdot}p_{2}\right)^{2}q^{\mu}\sum_{i,j\,i\ne j}\int\hat{d}^{4}l\frac{l{\cdot}(l-q)}{l^{2}(l-q)^{2}}\frac{\hat{\delta}(p_{j}{\cdot}l)}{(p_{i}{\cdot}l+i\epsilon)^{2}},\label{eq:I2}
\end{equation}
 Finally, we have the 4 point cut-box contribution 
\begin{equation}
\mathcal{I}_{3}^{\mu}=-2ig^4\left(Q_{1}Q_{2}p_{1}{\cdot}p_{2}\right)^{2}\int\hat{d}^{4}l\frac{l{\cdot}(l-q)l^{\mu}}{l^{2}(l-q)^{2}}\left[\hat{\delta}'(p_{1}{\cdot}l)\hat{\delta}(p_{2}{\cdot}l)-\hat{\delta}(p_{1}{\cdot}l)\hat{\delta}'(p_{2}{\cdot}l)\right].\label{eq:I3}
\end{equation}
By introducing  all these definitions we can check that the computation of $ {\cal T}^{\alpha}_{(1),\mathcal{R}_1}$ in 
 in \eqref{eq:R1-1}, toghether with the measure \eqref{eq:hbar0 measure}, can be rearrange to give exactly the NLO impulse \eqref{eq:NLO impulse}  plus an additional contribution coming from adding and subtracting the 4-pt cut-box diagram 
\begin{equation}
{\cal T}^{(1)\,\alpha}_{\mathcal{R}_1} = g^4\Delta p^{(1)\,\alpha}_{1} + {\rm [cut{-}box]^{(1)\,\alpha},}\label{eq:R1_final}
\end{equation}
with  the extra  contribution  ${\rm [cut{-}box]^{\mu}}$   given
by 
\begin{equation}
{\rm [cut{-}box]^{(1)\,\alpha}}=- g^4
     \left[p_{1,\beta}p_{1}^{[\beta}p_{2}^{\alpha]}-p_{2,\beta}p_{2}^{[\beta}p_{1}^{\alpha]}\right]\frac{\left(\Delta p^{(0)}_{1}\right)^{2}}{\mathcal{D}}\,,\label{eq:cut-box final}
\end{equation}
 The proof of this statement is lengthy  and we therefore  postpone it to be discussed  in   Appendix \ref{app:cut_box}.   
For the moment, let us notice that the first term of  eq. \eqref{eq:R1_final}
gives exactly the expected result from the second identity  \eqref{hier2}. Therefore, to conclude the proof we simply need to show that the remaining terms in \eqref{eq:1_index_moment_one_loop} together with \eqref{eq:cut-box final}, add up to zero. In fact, also in Appendix  \ref{app:cut_box} we will show that 
\begin{equation}\label{eq:cancel_r2_cb_c2}
   {\cal T}^{(1)\,\alpha}_{\mathcal{R}_2} + {\rm [cut{-}box]^{(1)\,\alpha}} = 0\,,
\end{equation}
whereas ${\cal T}^{(1)\,\alpha}_{\mathcal{C}_1}$ and ${\cal T}^{(1)\,\alpha}_{\mathcal{C}_2}$ evaluate to zero individually. 

In appendix \cref{app:NLO-IMPULSE} we  do the explicit computation of the NLO impulse \eqref{eq:NLO impulse} in the KMOC formalism, recovering the classical result \eqref{eq:NLO-impulse-justin}. Let us just mention here that the only two contributing  integrals are the triangle \eqref{eq:I1} and cut-box \eqref{eq:I3} integrals, giving the first and second terms in \eqref{eq:NLO-impulse-justin} respectively.  To make connection to the two-body potential calculation, these are the analog contributions appearance of the triangle, and the iterated tree-level amplitudes.

\subsubsection{Classical two-index \textit{moment} at 1-loop}\label{sec:r3}
The remaining task to complete the proof of identity \eqref{hier2} at 1-loop is to evaluate two-index \textit{moment}  \eqref{eq:two-index-moment-1-loop}. Similar to previous computation, we start  from its first term, given by  \eqref{eq:R3-1}, and  after inserting the measure \eqref{eq:hbar0 measure},  and the superclassical amplitude \eqref{eq:A4_super_classical}, we arrive at
\begin{equation}\label{eq:r3_start}
    {\cal T}^{(1)\,\alpha\,\beta}_{\mathcal{R}_3} =-\frac{1}{2}g^4\left(Q_{1}Q_{2}\,p_{1}{\cdot}p_{2}\right)^{2}   \int\hat{d}^{4}q\hat{d}^{4}l\prod_{i}\hat{\delta}(p_{i}{\cdot}q)\hat{\delta}(p_{i}{\cdot}l)e^{-ib{\cdot}q}\frac{q^{\alpha}q^{\beta}}{l^{2}(l-q)^{2}}\,.
\end{equation}
Next we can do our usual change of variables $q = l+\bar{q}$ 
\begin{equation}\label{eq:r3_start-1}
    {\cal T}^{(1)\,\alpha\,\beta}_{\mathcal{R}_3}=-\frac{1}{2}g^4\left(Q_{1}Q_{2}\,p_{1}{\cdot}p_{2}\right)^{2}  \int\hat{d}^{4}\bar{q}\hat{d}^{4}l\prod_{i}\hat{\delta}(p_{i}{\cdot}\bar{q})\hat{\delta}(p_{i}{\cdot}l)e^{-ib{\cdot}\bar{q}}e^{-ib{\cdot}l}\frac{\left(l^{\alpha}{+}\bar{q}^{\alpha}\right)\left(l^{\beta}{+}\bar{q}^{\beta}\right)}{l^{2}\bar{q}^{2}}\,.
\end{equation}
We  recognize the square of the leading order impulse \eqref{eq:LO_impulse_integral} coming from the crossed terms. On the other hand, the non-crossed terms give us the product of two integrals, one is them is the  usual IR-divergent
 integral  $I_{1}$ in $(\ref{eq:scalar integral I1})$, whereas the second one corresponds to the derivative 
of the leading order impulse  w.r.t. the impact parameter; notice  there is a factor of two for each case, which cancels the overall $1/2$ factor. That is 
\begin{equation}
   {\cal T}^{(1)\,\alpha\,\beta}_{\mathcal{R}_3} =  g^4{\color{black}} 
    \Delta p^{(0)\,\alpha}_{1}  \Delta p^{(0)\,\beta}_{1} 
   -g^4\left(Q_{1}Q_{2}\,p_{1}{\cdot}p_{2}\right)I_{1}\partial_{b^{\alpha}} \Delta p^{(0)\,\beta}_{1} \,.\label{eq:R3 interm2}
\end{equation}
The change of the  sign  for the first term comes from inserting a factor of $ i^2$ both, in the numerator and denominator, and absorb it  for the former, to complete the  the square of the leading order impulse. 
Using $(\ref{eq:scalar integral I1})$ and the derivative of the
leading order impulse 
\begin{equation}
\partial_{b^{\alpha}} \Delta p^{(0)\,\beta}_{1} =-\frac{Q_{1}Q_{2}p_{1}{\cdot}p_{2}}{2\pi\sqrt{\mathcal{D}}}\left(b^{2}\eta^{\alpha\beta}-2b^{\alpha}b^{\beta}\right)\frac{1}{b^{4}}\,,\label{eq:derivative lo-impulse wrt b}
\end{equation}
 and  drooping the term proportional to $\eta^{\alpha\beta}$ , using
the on-shell condition for the photon momentum and gauge invariance,
we finally arrive at 
\begin{equation}
   {\cal T}^{(1)\,\alpha\,\beta}_{\mathcal{R}_3} =g^4 \Delta p_{1}^{(0)\,\alpha}\Delta p_{1}^{(0)\,\beta}+\mathcal{J}_{3}^{(1)\,\alpha\beta}\,,\label{eq:R3 final}
\end{equation}
 where 
\begin{equation}
\mathcal{J}_{3}^{(1)\,\alpha\,\beta}= -2g^4\,\Delta p_{1}^{(0)\,\alpha}\Delta p_{1}^{(0)\,\beta}\ln\left(-\mu^{2}b^{2}\right)\,. \label{eq:J3}
\end{equation}

Similar to the previous subsection, to complete the proof of the second identity for the two-index \textit{moment} at 1-loop, we simple need to show that the second term in \eqref{eq:two-index-moment-1-loop} added to \eqref{eq:J3} evaluates to zero

\begin{equation}\label{eq:final_cancelations}
{\cal T}^{(1)\,\alpha\,\beta}_{\mathcal{C}_3} +\mathcal{J}_{3}^{(1)\,\alpha\,\beta} = 0\,.
\end{equation}
We leave the proof of this equation for Appendix \ref{app:cancelations_nlo_radiation}.

With this we have  concluded the proof of identity \eqref{hier2} at NLO in the perturbative expansion. 
Let us notice that the appearance of the square of the leading order impulse is a result of  $q$-expansion of the  Weinberg soft factor, iterated with the superclassical contributions from the box and cross box diagrams. However, remnants from the IR-divergent contributions as appearing in \eqref{eq:J3}, are nicely canceled by the $\mathcal{C}-$contribution to the radiated field, in analogy to the cancellation of IR divergent integrals from the EFT and full theory amplitudes matching \cite{Cheung:2018wkq,Bern:2019crd}. 

\section{Discussion}\label{sec:discusion}
The classical Soft theorems discovered by Weinberg, Saha, Sahoo and Sen provide us with universal and exact formulae for radiative electromagnetic and gravitational fields in a relativistic scattering. These formulae are for certain coefficients in the soft expansion of the radiative field. They only depend on the asymptotic kinematics of the incoming and outgoing bodies. Namely their mass, charge and momenta and are independent of the details of the scattering.\footnote{Of course given only the incoming data, the outgoing momenta are determined by the hard scattering, but the essence of classical soft theorem is in the observation that given the initial kinematics, the measurement of the momenta and charges of the outgoing particles in the asymptotic future can be used to determine the soft radiation.}
In fact, the theorems do not rely on validity of perturbation theory in the scattering region and provide us with formulae for memory effect, tail to the memory effect and so on produced in collisions of astrophysical objects \cite{Strominger:2014pwa,Laddha:2018vbn,Sahoo:2021ctw,Pasterski:2015tva}. 

In this paper, we have tried to analyse implications of classical soft theorems for KMOC formalism through which radiative field can be computed using on-shell techniques. As we have argued, classical soft theorems impose a tower of an  infinite hierarchy of constraints  on expectation values of a class of composite operators in the KMOC formalism. At leading order in the soft expansion, these operators are constructed from Monomials of Momentum operators. 

At leading order in perturbation theory, these constraints were verified in \cite{Bautista:2019tdr,Manu:2020zxl} at leading and sub-leading order in the soft expansion. In this paper,  we have  also verified them at NLO in the coupling and at leading order in the soft expansion in scalar QED with no higher derivative interactions.  We note that addition of other interactions will not change the structure of classical soft factor but will change the analytic expressions for the out going momenta in terms of incoming kinematics and impact parameter.  Verifying the sub-leading soft constraints at higher orders in the perturbative expansion requires a deeper investigation into the integration regions involving the loop momenta. This analysis is under progress and will be reported elsewhere.

At NLO, the verification of the leading soft constraint is analogous to the EFT and full theory amplitudes matching procedure for the computation of the 2PM two-body potential \cite{Cheung:2018wkq,Bern:2019crd}. A difference between  the two computations is in the treatment of  super-classical terms. In the soft constraints derived from KMOC formalism, the IR divergent terms  cancel by the addition of the $\mathcal{C}$-contributions to the radiative field \eqref{c-ker1}, in contrast to the   matching procedure. We have also seen that the powers of the leading order impulse were the analogs to the iterated tree-level amplitudes appearing in the 2PM potential. Furthermore,  contribution to the NLO impulse coming from the triangle  and cut-box integrals have the respective counterpart in the 2PM potential. Viewed in this light, the classical soft theorems impose constraints in the elastic dynamics of the two-body problem. Indeed, once the frequency of the radiated photon (graviton) is fixed, soft-theorems become an statement on the elastic sector\footnote{At sufficient higher orders in perturbation theory, radiation reaction effects eventually get manifest in the elastic sector  and need to be incorporated in the definition of the linear impulse  in order for constraints (\ref{hier1}-\ref{hier2}) to be satisfied. We thank the referees of this paper for pointing this out.  }. At 3PM for instance, the appearance of iterative 1-loop and tree-level contributions  to the potential \cite{Bern:2019crd,Bjerrum-Bohr:2021din, Kalin:2020fhe}, will be the analogs of products of the form $\Delta p^{(1)}{\cdot}\Delta p^{(0)}$, appearing at two loops in \eqref{two}, in addition to the cubic appearance of the tree-level amplitude, which will be the analog of $(\Delta p^{(0)})^3$, and analogously for the 4PM result \cite{Bern:2021dqo,Dlapa:2021npj}

In this paper we have solely focused on soft  electro-magnetic radiation. We believe that the leading soft constraints can be generalised to gravitational interactions directly at NLO. Beyond NLO order, classical soft graviton factor will receive contribution from finite energy gravitational flux. On the other hand if we take classical limit after applying Weinberg soft theorem inside the radiation kernel, the result will be once again turn out to be in terms of monomials of linear impulses.  We believe that this result once again should be equated to the contribution to the classical soft graviton factor only from outgoing massive particles. However this remains to be shown. As KMOC naturally takes into account the dissipative effects in computation of linear impulse, we expect this procedure to be consistent. \footnote{We thank Ashoke Sen for discussion on this issue.} \footnote{The generalisation of the sub-leading soft constraints may be even more subtle as the classical log soft factor in gravity has an additional contribution effect of space-time curvature on soft radiation. These terms may not simply arise from sub-leading soft graviton theorem for the integrands \cite{Sahoo:2018lxl}.}

It will be interesting to prove the leading and sub-leading soft constraints within KMOC formalism  for perturbative scattering with large impact parameter. Universality of classical soft theorems imply that the proof is likely to involve ideas along the lines of the classical proof in  \cite{Saha:2019tub}, in which it was only assumed that the interactions outside a ``hard scattering region" (which can be parametrized as a space-time region bounded in spatial and temporal directions by $\pm\, t_{0}$ for some sufficiently large $t_{0}$) are simply the Coulombic interactions. However formulating the quantum dynamics in this fashion may require use of the time-ordered perturbation theory \cite{Sterman1993} which has in fact also been adopted to hard-soft factorisation in the seminal paper by Schwartz and Hannesdottir \cite{Hannesdottir:2019opa}.

\acknowledgments

 We are  grateful to Laurent Freidel, Alfredo Guevara, Athira P V, Akavoor Manu, Partha Paul, Ashoke Sen,  Justin Vines and Yong Zhang for enlightening discussions.  YFB  acknowledges financial support by the Natural Sciences and Engineering Research Council of Canada (NSERC). Research at Perimeter Institute is supported in part by the Government of Canada through the Department of Innovation, Science and Economic Development Canada and by the Province of Ontario through the Ministry of Economic Development, Job Creation and Trade.

\appendix
\section{Computational details NLO radiation}\label{app:radiation_NLO}
In this appendix we walk through the computational details of several integrals given in section \cref{sec:1-loop_radiation}
\subsection{Cancellations in the classical one-index \textit{moment} at 1-loop
}\label{app:cut_box}
In this section we fill in  the computational details for the cancellations announced by the end of  \cref{sec:one-index-1-loop}.
\begin{itemize}
    \item \textit{Proof of eq. \eqref{eq:cut-box final}}
\end{itemize}
Let us begin  by walking through the proof of the stamen of equation \eqref{eq:cut-box final}. We start from 
\begin{equation}\label{eq:cut_bot_start}
  {\rm [cut{-}box]^{(1)\,\alpha}}=\frac{1}{4}\int\hat{d}^{4}q\prod_{i}\hat{\delta}(p_{i}{\cdot}q)e^{-ib{\cdot}q}\left(\mathcal{I}_{3}^{\alpha}-q^\alpha Z=U^{\alpha}\right)\,,
\end{equation}
where  the integrand has the explicit form
\begin{equation}\label{eq:u_integrand_in_cut-box}
U^{\alpha}=-2g^4\left(Q_{1}Q_{2}p_{1}{\cdot}p_{2}\right)^{2}\int\frac{\hat{d}^{4}l}{l^{2}(l{-}q)^{2}}\left[\hat{\delta}'(p_{1}{\cdot}l)\hat{\delta}(p_{2}{\cdot}l){-}\hat{\delta}(p_{1}{\cdot}l)\hat{\delta}'(p_{2}{\cdot}l)\right]\left[l{\cdot}(l-q)l^{\alpha}{+}\frac{q^{\alpha}}{2}(2l{\cdot}q{-}l^{2})\right]\,.
\end{equation}
 To evaluate this integral  we can  expand the momentum $l$ in an analogous way to the $q$ momentum in (\ref{eq:momentum-expansion-q} - \ref{eq:alphai_factors}), with say $\alpha_i\to \beta_i$, and $x_i\to y_i= p_i\cdot l$. The resulting  integrand  takes the form   \eqref{eq:general_integral}, and therefore we can evaluate the time and longitudinal components using integrating by parts one time \eqref{eq:general_ibp}. That is, we can write 
 \begin{equation}
     U^\alpha = \frac{1}{\sqrt{\mathcal{D}}}\int  \hat{d}^{2}l_{\perp}\hat{d}y_1\hat{d}y_2\left[ \hat{\delta}^{(1)}(y_1)\hat{\delta}^{(0)}(y_2) - \hat{\delta}^{(0)}(y_1)\hat{\delta}^{(1)}(y_2)\right]  f_u^\alpha(y_1,y_2,l_\perp,\sigma),
 \end{equation}
 with the  identification of the  integrand function   
\begin{align}
f_u^\alpha(y_1,y_2,l_\perp,\sigma)=-2g^4\left(Q_{1}Q_{2}p_{1}{\cdot}p_{2}\right)^{2}\Bigg[ & \frac{(\beta_{2}p_{1}{+}\beta_{1}p_{2}+l_{\perp})^{\alpha}\left((\beta_{2}p_{1}{+}\beta_{1}p_{2})^{2}{+}l_{\perp}^{2}{-}l_{\perp}{\cdot}q_{\perp}\right)}{\left((\beta_{2}p_{1}{+}\beta_{1}p_{2})^{2}{+}l_{\perp}^{2}\right)\left((\beta_{2}p_{1}{+}\beta_{1}p_{2})^{2}{+}(l_{\perp}{-}q_{\perp})^{2}\right)}\nonumber \\
 & {-}\frac{\left((\beta_{2}p_{1}{+}\beta_{1}p_{2})^{2}{+}l_{\perp}^{2}-2l_{\perp}{\cdot}q_{\perp}\right)\frac{q_{\perp}^{\alpha}}{2}}{\left((\beta_{2}p_{1}{+}\beta_{1}p_{2})^{2}{+}l_{\perp}^{2}\right)\left((\beta_{2}p_{1}{+}\beta_{1}p_{2})^{2}{+}(l_{\perp}{-}q_{\perp})^{2}\right)}\Bigg]\,.\label{eq:fu1}
\end{align}
where in addition to the $l$-expansion, we have used the expansion for the $q$-momentum \eqref{eq:momentum-expansion-q}, and set $x_i\to0$ using the support of the delta function $\hat{\delta}(p_i{\cdot}q)$ in \eqref{eq:cut_bot_start}. Next, to use \eqref{eq:general_ibp} after integration by parts we need to evaluate the derivatives of the form 
\begin{equation}\label{eq:derf}
\frac{\partial}{\partial y_{i}}f_{u,ij}^{\alpha}\Bigg|_{y_{i}=y_{j}=0}=\frac{4g^4}{\mathcal{D}}\left(Q_{i}Q_{j}p_{i}{\cdot}p_{j}\right)^{2}p_{j,\beta}p_{j}^{[\beta}p_{i}^{\alpha]}\frac{l_{\perp}{\cdot}(l_{\perp}-q_{\perp})}{l_{\perp}^{2}(l_{\perp}-q_{\perp})^{2}}\,,
\end{equation}
where one can check that only the first line of \eqref{eq:fu1} contributed to \eqref{eq:derf}.
With all the tools at hand, it is then direct to show that the integral \eqref{eq:cut_bot_start} simplifies to
\begin{equation}
{\rm [cut{-}box]^{(1)\,\alpha}}=g^4\frac{\left(Q_{1}Q_{2}p_{1}{\cdot}p_{2}\right)^{2}}{\mathcal{D}^2}\left[p_{2,\beta}p_{2}^{[\beta}p_{1}^{\alpha]}{-}p_{1,\beta}p_{1}^{[\beta}p_{2}^{\alpha]}\right]\int\hat{d}^{2}q_{\perp}\hat{d}^{2}l_{\perp}e^{-ib{\cdot}q_{\perp}}\frac{l_{\perp}{\cdot}(l_{\perp}{-}q_{\perp})}{l_{\perp}^{2}(l_{\perp}{-}q_{\perp})^{2}}.\label{eq:cut-box int2}
\end{equation}
Next we do the usual change of variables $q_{\perp}=\bar{q}_{\perp}+l_\perp,$
so that 
\begin{equation}
{\rm [cut{-}box]^{(1)\,\alpha}}=\frac{g^4}{\mathcal{D}}\left[p_{2,\beta}p_{2}^{[\beta}p_{1}^{\alpha]}{-}p_{1,\beta}p_{1}^{[\beta}p_{2}^{\alpha]}\right]\left[\frac{Q_{1}Q_{2}p_{1}{\cdot}p_{2}}{\sqrt{\mathcal{D}}}\int\hat{d}^{2}q_{\perp}\hat{d}^{2}l_{\perp}e^{-ib{\cdot}q_{\perp}}i\frac{\bar{q}_{\perp}}{\bar{q}_{\perp}^{2}}\right]^{2}.\label{eq:cut-box int3}
\end{equation}
in the big bracket we recognize the Leading order impulse $(\ref{eq:LO_impulse_interm})$,
which in turn allow us to recover the announced result $(\ref{eq:cut-box final}).$
\begin{equation}
{\rm [cut{-}box]^{(1)\,\alpha}}= - g^4
     \left[p_{1,\beta}p_{1}^{[\beta}p_{2}^{\alpha]}-p_{2,\beta}p_{2}^{[\beta}p_{1}^{\alpha]}\right]\frac{\left(\Delta p^{(0)}_{1}\right)^{2}}{\mathcal{D}}\,,\label{eq:cut-box final_app}
\end{equation}
\begin{itemize}
    \item \textit{Proof of eq. \eqref{eq:cancel_r2_cb_c2}}
\end{itemize}
Next we move to prove the cancellation \eqref{eq:cancel_r2_cb_c2}. For that we still need to compute $ {\cal T}^{\alpha}_{(1),\mathcal{R}_2}$ starting from \eqref{eq:R2-1}, and using  \eqref{eq:hbar1measure} for the integral  measure, and \eqref{eq:4pt_LO} for the 4 point amplitude. We get

\begin{equation}\begin{split}
     {\cal T}^{(1)\,\alpha}_{\mathcal{R}_2} =
\frac{1}{4}g^4\left(Q_{1}Q_{2}\,p_{1}{\cdot}p_{2}\right)^{2}&\int\hat{d}^{4}l\prod_{i}\hat{\delta}(p_{i}{\cdot}l)\hat{d}^{4}q e^{-ib{\cdot}q}\,\\
&\times\left[\hat{\delta}'(p_{1}{\cdot}q)\hat{\delta}(p_{2}{\cdot}q){-}\hat{\delta}(p_{1}{\cdot}q)\hat{\delta}'(p_{2}{\cdot}q)\right]\frac{q^{2}q^{\alpha}}{l^{2}(l-q)^{2}}\,,
\end{split}
\end{equation}
where we have  used $\delta'(2p_{i}{\cdot}q)=\frac{1}{4}\delta'(p_{i}{\cdot}q)$.
To proceed in the calculation, we follow the philosophy of the previous subsection for the computation of integrals involving derivatives of the Dirac delta function, i.e. using integration by parts. Doing the change of variables (\ref{eq:momentum-expansion-q}) (and the analogous change for  $q\to l$ and $x_i\to y_i$ ), and evaluating the integrals in $(p_{i}{\cdot}l) =  y_i $, using the corresponding delta function, we arrive at 
\begin{equation}\label{eq:R2-int}
    {\cal T}^{(1)\,\alpha}_{\mathcal{R}_2}= \frac{1}{\mathcal{D}}\int\hat{d}^{2}q_{\perp}\frac{\hat{d}^{2}l_{\perp}}{l_{\perp}^{2}}\prod_{i}\hat{d}x_i\left[\hat{\delta}'(x_i)\hat{\delta}(x_2){-}\hat{\delta}(x_1)\hat{\delta}'(x_2)\right]Y^{\alpha}\,,
\end{equation}
where we have defined
\begin{equation}
Y^{\alpha}=\frac{1}{4}g^4\left(Q_{1}Q_{2}\,p_{1}{\cdot}p_{2}\right)^{2}e^{-iq_{\perp}{\cdot}b}\frac{((\alpha_{2}p_{1}{+}\alpha_{1}p_{2})^{2}+q_{\perp}^{2})(\alpha_{2}p_{1}{+}\alpha_{1}p_{2}{+}q_{\perp})^{\alpha}}{(\alpha_{2}p_{1}{+}\alpha_{1}p_{2})^{2}{+}(l_{\perp}-q_{\perp})^{2}}\,,\label{eq:Y function}
\end{equation}
recalling that    $\alpha_i$ are  function of $x_i$. We then get an integral of the form \eqref{eq:general_integral}. Using 
\begin{equation}
\frac{\partial}{\partial x_i}Y_{ij}^{\alpha}\Bigg|_{x_1=x_2=0}=-\frac{1}{2}g^4\left(Q_{1}Q_{2}\,p_{1}{\cdot}p_{2}\right)^{2}\frac{q_{\perp}^{2}}{\mathcal{D}(l_{\perp}-q_{\perp})^{2}}p_{j,\nu}p_{j}^{[\nu}p_{i}^{\alpha]}e^{-iq_{\perp}{\cdot}b}\label{eq:derivative Y},
\end{equation}
for doing the integration  by parts procedure, we can write \eqref{eq:R2-int} as follows
\begin{equation}
     {\cal T}^{(1)\,\alpha}_{\mathcal{R}_2} ={-}\frac{1}{2}\frac{g^4\left(Q_{1}Q_{2}\,p_{1}{\cdot}p_{2}\right)^{2}}{\mathcal{D}^2}\left(p_{1,\beta}p_{1}^{[\beta}p_{2}^{\alpha]}{-}p_{2,\beta}p_{2}^{[\beta}p_{1}^{\alpha]}\right)
\left[\int\hat{d}^{2}q_{\perp}\hat{d}^{2}l_{\perp}\frac{q_{\perp}^{2}e^{-iq_{\perp}{\cdot}b}}{l_{\perp}^{2}(l_{\perp}-q_{\perp})^{2}}{=}\mathcal{J}_{2}\right]\,.\label{eq:R2 interm 2}
\end{equation}
Let us now  evaluate the integral $\mathcal{J}_{2}$ in the square brackets.  Doing the usual shift
$q_{\perp}=\bar{q}_{\perp}+l_{\perp},$ so that
\begin{equation}
\mathcal{J}_{2}=\int\hat{d}^{2}\bar{q}_{\perp}\hat{d}^{2}l_{\perp}\frac{(\bar{q}_{\perp}^{2}+l_{\perp}^{2}+2l_{\perp}{\cdot}\bar{q}_{\perp})e^{-i\bar{q}_{\perp}{\cdot}b}e^{-il_{\perp}{\cdot}b}}{l_{\perp}^{2}\bar{q}_{\perp}^{2}}\,.\label{eq:J2 definition}
\end{equation}
 From the crossed terms we identify the integral representation for the  leading order impulse, whereas the remaining terms are contact integrals which we drop assuming $b\ne0$. We finally get
\begin{equation}
\mathcal{J}_{2}=-2\frac{\left(\Delta p^{(0)}\right)^{2}\mathcal{D}}{g^4\left(Q_{1}Q_{2}\,p_{1}{\cdot}p_{2}\right)^{2}}\,,\label{eq:J2 final}
\end{equation}
which can be replace back  into \eqref{eq:R2 interm 2} to finally give
\begin{equation}
 {\cal T}^{(1)\,\alpha}_{\mathcal{R}_2} = g^4 \left[p_{1,\beta}p_{1}^{[\beta}p_{2}^{\alpha]}{-}p_{2,\beta}p_{2}^{[\beta}p_{1}^{\alpha]}\right]\frac{\left(\Delta p^{(0)}\right)^{2}}{\mathcal{D}}\,.\label{eq:R2 final}
\end{equation}
We note that this simply gives  ${\cal T}^{(1)\,\alpha}_{\mathcal{R}_2} =- {\rm [cut{-}box]^{(1)\,\alpha}}$, and therefore this concludes the proof of  \eqref{eq:cancel_r2_cb_c2}.
\begin{itemize}
    \item \textit{Vanishing of ${\cal T}^{(1)\,\alpha}_{\mathcal{C}_1}$}
\end{itemize}
This is a very simple proof since this term give us a contact integral. Our  starting point is  the definition \eqref{eq:C1 definition-2}, and using \eqref{eq:hbar0 measure}  and \eqref{eq:hbar1measure} for the integral measure in $q$ and $l$ respectively, and \eqref{eq:4pt_LO} for the 4 point amplitude, we get  
\begin{equation}
\begin{split}
{\cal T}^{(1)\,\alpha}_{\mathcal{C}_1}  =  16 g^4\left(Q_{1}Q_{2}p_{1}{\cdot}p_{2}\right)^{2}&\int\hat{d}^{4}q\hat{d}^{4}l e^{-ib{\cdot}q}\prod_{i}\hat{\delta}(2p_{i}{\cdot}q)\\
 &\times \left[\hat{\delta}'(2p_{1}{\cdot}l)\hat{\delta}(2p_{2}{\cdot}l){-}\hat{\delta}(2p_{1}{\cdot}l)\hat{\delta}'(2p_{2}{\cdot}l)\right]
 \frac{\cancel{l^{2}}l^{\alpha}}{\cancel{l^{2}}(q-l)^{2}}
\end{split}\,,
\end{equation}
which gives us indeed a  contact integral for the $l$-variable, unimportant for long classical scattering, since $b\ne0$. 
\begin{itemize}
    \item \textit{Vanishing of ${\cal T}^{(1)\,\alpha}_{\mathcal{C}_2}$}
\end{itemize}
Here we carry out the final piece of the computation for the one-index \textit{moment}. 
As usual, we start from the definition \eqref{eq:C2 definition-2}, and use \eqref{eq:hbar1measure} and \eqref{eq:hbar0 measure} for the integral measure in $q$ and $l$ respectively,  and \eqref{eq:4pt_LO} for the 4 point amplitude. This gives

\begin{equation}
{\cal T}^{(1)\,\alpha}_{\mathcal{C}_2}= \int\hat{d}^{4}q\hat{d}^{4}l\prod_{i}\hat{\delta}(p_{i}{\cdot}l)\left[\hat{\delta}'(p_{1}{\cdot}q)\hat{\delta}(p_{2}{\cdot}q){-}\hat{\delta}(p_{1}{\cdot}q)\hat{\delta}'(p_{2}{\cdot}q)\right]X^{\alpha}\,,
\end{equation}
where we have defined 
\begin{equation}
X^{\alpha}=-\frac{1}{2}g^4\left(Q_{1}Q_{2}p_{1}{\cdot}p_{2}\right)^{2}\frac{q^{2}l^{\alpha}}{l^{2}(q{-}l)^{2}}e^{-ib{\cdot}q}\,.
\end{equation}
Showing that this integral gives zero contribution is a straightforward task. We do the usual change of variables 
 \eqref{eq:momentum-expansion-q}, and doing the integrals in $y_i$ using the delta functions. However, since we will use integration by parts, we need to evaluate the derivative of $X^{\alpha}$ w.r.t. $x_1$ or $x_2$, which  after evaluating $x_1=x_2=0$, vanish identically. We therefore conclude that  ${\cal T}^{\alpha}_{(1),\mathcal{C}_2}=0$, as announced in \cref{sec:one-index-1-loop}

\subsection{Cancellations in the classical two-index \textit{moment} at 1-loop}\label{app:cancelations_nlo_radiation}
In the final part of this appendix we proof the cancellations announced  in 
\eqref{eq:final_cancelations}.  Recall we already obtained $\mathcal{J}_3$ in \eqref{eq:J3}. All that is left is to compute explicitly ${\cal T}^{\alpha\,\beta}_{(1),\mathcal{C}_3} $. As usual we start  from the definition \eqref{eq:C3 definition-2}. Use \eqref{eq:hbar0 measure} for the integral measure in both $q$ and $l$ variables, and \eqref{eq:4pt_LO} for the 4 point  amplitude, to get
\begin{equation}
{\cal T}^{(1)\,\alpha\,\beta}_{\mathcal{C}_3}= g^4\left(Q_{1}Q_{2}p_{1}{\cdot}p_{2}\right)^{2}\int\hat{d}^{4}q\hat{d}^{4}l\prod_{i}\hat{\delta}(p_{i}{\cdot}q)\hat{\delta}(p_{i}{\cdot}l)\frac{l^{\alpha}l^{\beta}}{l^{2}(q{-}l)^{2}}e^{-ib{\cdot}q}\,.\label{C3 start}
\end{equation}
 Doing our usual  shift $q=\bar{q}+l,$ allows us to factorize out the  IR-divergent integral - for the $l$ variable - \eqref{eq:scalar integral I1}. This becomes
\begin{equation}
{\cal T}^{(1)\,\alpha\,\beta}_{\mathcal{C}_3} = g^4\left(Q_{1}Q_{2}p_{1}{\cdot}p_{2}\right)^{2}I_{1} \int\hat{d}^{4}\bar{q}\hat{\delta}(p_{1}{\cdot}\bar{q})\hat{\delta}(p_{2}{\cdot}\bar{q})\frac{l^{\alpha}l^{\beta}}{\bar{q}^{2}}e^{-ib{\cdot}\bar{q}}\,,\label{eq:C3 interm 2}
\end{equation}
 which can be further rewritten as 
\begin{equation}
{\cal T}^{(1)\,\alpha\,\beta}_{\mathcal{C}_3}  =g^4\left(Q_{1}Q_{2}p_{1}{\cdot}p_{2}\right)I_{1} \partial_{b^{\alpha}}\Delta p^{(0)\,\beta}_{1}\,.\label{eq:C3 interm 4}
\end{equation}
The computation of  the derivative of the leading order impulse   w.r.t. the impact parameter was given in  \eqref{eq:derivative lo-impulse wrt b}. Using it leads to 
\begin{equation}
    {\cal T}^{\alpha\,\beta}_{(1),\mathcal{C}_3} =  2 g^4 \Delta p^{(0)\,\alpha}_{1} \Delta p^{(0)\,\beta}_{1}\ln\left(-\mu^{2}b^{2}\right),
\end{equation}
which is nothing but $\mathcal{J}_3^{(1)\,\alpha\,\beta}$ as given in 
\eqref{eq:J3} but with opposite sign. Thus we simply conclude that 
$
{\cal T}^{(1)\,\alpha\,\beta}_{\mathcal{C}_3} +\mathcal{J}_{3}^{(1)\,\alpha\,\beta} = 0\,,$ as required from \eqref{eq:final_cancelations}.

\section{Electro-magnetic impulse at NLO from KMOC}\label{app:NLO-IMPULSE}

In this appendix we show how to obtain the classical result for the NLO impulse \eqref{eq:NLO-impulse-justin} from the amplitudes integrals \eqref{eq:NLO impulse} resulting from the KMOC formalism. Let us start by evaluating the last term in  \eqref{eq:NLO impulse} using \eqref{eq:I3},

\begin{equation}
    \tilde{\mathcal{I}}_3^\mu= \frac{g^4}{2}(Q_1Q_2p_q{\cdot}p_2)^2 \int \hat{d}^4q\hat{d}^4l \hat{\delta}(p_1{\cdot}q)\hat{\delta}(p_2{\cdot}q) \frac{l{\cdot}(l{-}q)}{l^2(l{-}q)^2}l^\mu e^{-iq{\cdot}b}[\hat{\delta}'(p_1{\cdot}l)\hat{\delta}(p_2{\cdot}l){-}\hat{\delta}(p_1{\cdot}l)\hat{\delta}'(p_2{\cdot}l)],
\end{equation}
where the tilde over $ \mathcal{I}_3$ indicates inclusion of the $q$ integration. 
This integral is analog to the cut-box integral \eqref{eq:cut_bot_start}, where indeed only the first term of the last square bracket in \eqref{eq:u_integrand_in_cut-box} contributes. The final result will simply be

\begin{equation}
     \tilde{\mathcal{I}}_3^\mu=g^4 \frac{(Q_1Q_2p_1{\cdot}p_2)^2}{8\pi^2\mathcal{D}^2|b|^2}\left[\left(m_1^2+p_1{\cdot}p_2 \right)p_2^\mu -\left(m_2^2+p_1{\cdot}p_2 \right)p_1^\mu \right].
\end{equation}
This indeed provides us with the second term in the classical impulse \eqref{eq:NLO-impulse-justin}, with the  contributions only from the cut-box diagrams. 

Let us now move to the computation of the triangle diagrams which corresponds to  the first term in  \eqref{eq:NLO impulse}. Using \eqref{eq:I1}, we have

\begin{equation}
    \tilde{\mathcal{I}}_1^\mu =\frac{ig^4}{2}(Q_1Q_2)^2\int \hat{d}^4q\hat{d}^4l\hat{\delta}(p_1{\cdot}q)\hat{\delta}(p_2{\cdot}q) \frac{q^\mu}{l^2(l-q)^2}  e^{-iq{\cdot}b}\left[m_1^2 \hat{\delta}(p_1{\cdot}l) +m_2^2 \hat{\delta}(p_2{\cdot}l) \right] \,.
\end{equation}
We can do the integral in $l^0$ by going to the rest frame of particle $1$ (or $2$) then getting $\hat{\delta}(l^0)$. Notice that we can  also set $q^0=0$ by using one of the on-shell delta functions in $q$. With this in mind it follows
\begin{equation}
    \tilde{\mathcal{I}}_1^\mu =\frac{ig^4}{2}(Q_1Q_2)^2(m_1 +m_2 )\int \hat{d}^4q\hat{d}^3\vec{l}\hat{\delta}(p_1{\cdot}q)\hat{\delta}(p_2{\cdot}q) \frac{q^\mu}{\vec{l}^2(\vec{l}-\vec{q})^2}  e^{-iq{\cdot}b}\,.
\end{equation}
The integral in $\hat{d}^3\vec{l}$  is easy to evaluate using Schwinger parameters, see for instance eq. $(7.9)$ in \cite{Bern:2019crd} . Using those results we  get 
\begin{equation}
    \tilde{\mathcal{I}}_1^\mu =\frac{ig^4}{16\sqrt{D}}(Q_1Q_2)^2(m_1 +m_2 )\int \hat{d}^2q_\perp \frac{q_\perp^\mu}{q_\perp}  e^{-iq_\perp{\cdot}b},
\end{equation}
where we have further evaluated  two of the $\hat{d}^4q$ integrals using the expansions for the momenta \eqref{eq:expansion_momenta}. Evaluation of the remaining integral can be done in polar coordinates, upon trading  $q_\perp^\mu$ in the numerators by a derivative w.r.t. the impact parameter. The final answer will be 

\begin{equation}
      \tilde{\mathcal{I}}_1^\mu =-\frac{g^4}{32\pi}(m_1+m_2)\frac{(Q_1Q_2)^2}{\sqrt{D}}\frac{b^\mu}{|b|^3}.
\end{equation}
This in turn recovers the first term in the classical impulse \eqref{eq:NLO-impulse-justin}, with contributions coming only from the two triangle diagrams. 

Then, the final task is to show that the box and cross-box diagrams from  integral \eqref{eq:I2} gives vanishing contribution.  We can see this by first  dropping the term proportional to $l^2 $ in the numerator since it give rise  to non local contributions. Next, using the same philosophy of \cite{Kalin:2020mvi}, we can write $2 l\cdot q=l^2+q^2-(l-q)^2$, and discarding again non local contributions; the integral  \eqref{eq:I2} becomes 

\begin{equation}
\mathcal{I}_{2}^{\mu}=-g^4\left(Q_{1}Q_{2}p_{1}{\cdot}p_{2}\right)^{2}q^{\mu}\sum_{i,j\,i\ne j}q^2\int\hat{d}^{4}l\frac{\hat{\delta}(p_{j}{\cdot}l)}{l^{2}(l-q)^{2}(p_{i}{\cdot}l+i\epsilon)^{2}},\label{eq:I2n}
\end{equation}
using  the fact that at NLO no net four-momentum is radiated, radiation poles do not contribute to the integral, we can choose a contour in the opposite half of the plane were  \eqref{eq:I2} has the double poles $(p{\cdot}l+i\epsilon)^2$, and then getting a vanishing integral.  Indeed this was done for the gravitational case in   \cite{Kalin:2020mvi} eq. (4.26).
In conclusion, the NLO electro-magnetic impulse 

\begin{equation}
\Delta p_1^{(1)\,\mu} =  \tilde{\mathcal{I}}_1^\mu+ \tilde{\mathcal{I}}_3^\mu = -\frac{g^{4}}{32\pi^2|b|^{3}}\frac{(Q_{1}Q_{2})^2}{\mathcal{D}}\left[\pi\sqrt{\mathcal{D}}(m_{1}+m_{2}) b^{\mu}+4\frac{(p_{1}{\cdot}p_{2})^2(p_{1}{+}p_{2})^{2}|b|}{\mathcal{D}}p^{\mu}\right]\,,
\end{equation}
where $p^\mu$ was defined in \eqref{eq:p-def_justin}, therefore recovering the result of \cite{Saketh:2021sri}.

\bibliographystyle{JHEP}
\bibliography{references}

\providecommand{\href}[2]{#2}\begingroup\raggedright\begin{thebibliography}{10}

\bibitem{Laddha:2018rle}
A.~Laddha and A.~Sen, {\it {Gravity Waves from Soft Theorem in General
  Dimensions}},  {\em JHEP} {\bf 09} (2018) 105,
  [\href{http://arxiv.org/abs/1801.07719}{{\tt arXiv:1801.07719}}].

\bibitem{Laddha:2018myi}
A.~Laddha and A.~Sen, {\it {Logarithmic Terms in the Soft Expansion in Four
  Dimensions}},  {\em JHEP} {\bf 10} (2018) 056,
  [\href{http://arxiv.org/abs/1804.09193}{{\tt arXiv:1804.09193}}].

\bibitem{Sahoo:2018lxl}
B.~Sahoo and A.~Sen, {\it {Classical and Quantum Results on Logarithmic Terms
  in the Soft Theorem in Four Dimensions}},  {\em JHEP} {\bf 02} (2019) 086,
  [\href{http://arxiv.org/abs/1808.03288}{{\tt arXiv:1808.03288}}].

\bibitem{Laddha:2019yaj}
A.~Laddha and A.~Sen, {\it {Classical proof of the classical soft graviton
  theorem in D\ensuremath{>}4}},  {\em Phys. Rev. D} {\bf 101} (2020), no.~8
  084011, [\href{http://arxiv.org/abs/1906.08288}{{\tt arXiv:1906.08288}}].

\bibitem{Saha:2019tub}
A.~P. Saha, B.~Sahoo, and A.~Sen, {\it {Proof of the classical soft graviton
  theorem in $D$ = 4}},  {\em JHEP} {\bf 06} (2020) 153,
  [\href{http://arxiv.org/abs/1912.06413}{{\tt arXiv:1912.06413}}].

\bibitem{Sahoo:2020ryf}
B.~Sahoo, {\it {Classical Sub-subleading Soft Photon and Soft Graviton Theorems
  in Four Spacetime Dimensions}},  {\em JHEP} {\bf 12} (2020) 070,
  [\href{http://arxiv.org/abs/2008.04376}{{\tt arXiv:2008.04376}}].

\bibitem{Ghosh:2021hsk}
D.~Ghosh and B.~Sahoo, {\it {Spin Dependent Gravitational Tail Memory in
  $D=4$}},  \href{http://arxiv.org/abs/2106.10741}{{\tt arXiv:2106.10741}}.

\bibitem{PhysRev.140.B516}
S.~Weinberg, {\it Infrared photons and gravitons},  {\em Phys. Rev.} {\bf 140}
  (Oct, 1965) B516--B524.

\bibitem{Freidel:2021qpz}
L.~Freidel and D.~Pranzetti, {\it {Gravity from symmetry: Duality and impulsive
  waves}},  \href{http://arxiv.org/abs/2109.06342}{{\tt arXiv:2109.06342}}.

\bibitem{Kosower:2018adc}
D.~A. Kosower, B.~Maybee, and D.~O'Connell, {\it {Amplitudes, Observables, and
  Classical Scattering}},  {\em JHEP} {\bf 02} (2019) 137,
  [\href{http://arxiv.org/abs/1811.10950}{{\tt arXiv:1811.10950}}].

\bibitem{Maybee:2019jus}
B.~Maybee, D.~O'Connell, and J.~Vines, {\it {Observables and amplitudes for
  spinning particles and black holes}},  {\em JHEP} {\bf 12} (2019) 156,
  [\href{http://arxiv.org/abs/1906.09260}{{\tt arXiv:1906.09260}}].

\bibitem{Guevara:2019fsj}
A.~Guevara, A.~Ochirov, and J.~Vines, {\it {Black-hole scattering with general
  spin directions from minimal-coupling amplitudes}},  {\em Phys. Rev. D} {\bf
  100} (2019), no.~10 104024, [\href{http://arxiv.org/abs/1906.10071}{{\tt
  arXiv:1906.10071}}].

\bibitem{Cristofoli:2021vyo}
A.~Cristofoli, R.~Gonzo, D.~A. Kosower, and D.~O'Connell, {\it {Waveforms from
  Amplitudes}},  \href{http://arxiv.org/abs/2107.10193}{{\tt
  arXiv:2107.10193}}.

\bibitem{Aoude:2021oqj}
R.~Aoude and A.~Ochirov, {\it {Classical observables from coherent-spin
  amplitudes}},  {\em JHEP} {\bf 10} (2021) 008,
  [\href{http://arxiv.org/abs/2108.01649}{{\tt arXiv:2108.01649}}].

\bibitem{Herrmann:2021lqe}
E.~Herrmann, J.~Parra-Martinez, M.~S. Ruf, and M.~Zeng, {\it {Gravitational
  Bremsstrahlung from Reverse Unitarity}},  {\em Phys. Rev. Lett.} {\bf 126}
  (2021), no.~20 201602, [\href{http://arxiv.org/abs/2101.07255}{{\tt
  arXiv:2101.07255}}].

\bibitem{Cristofoli:2021jas}
A.~Cristofoli, R.~Gonzo, N.~Moynihan, D.~O'Connell, A.~Ross, M.~Sergola, and
  C.~D. White, {\it {The Uncertainty Principle and Classical Amplitudes}},
  \href{http://arxiv.org/abs/2112.07556}{{\tt arXiv:2112.07556}}.

\bibitem{Bautista:2019tdr}
Y.~F. Bautista and A.~Guevara, {\it {From Scattering Amplitudes to Classical
  Physics: Universality, Double Copy and Soft Theorems}},
  \href{http://arxiv.org/abs/1903.12419}{{\tt arXiv:1903.12419}}.

\bibitem{Manu:2020zxl}
A.~Manu, D.~Ghosh, A.~Laddha, and P.~V. Athira, {\it {Soft radiation from
  scattering amplitudes revisited}},  {\em JHEP} {\bf 05} (2021) 056,
  [\href{http://arxiv.org/abs/2007.02077}{{\tt arXiv:2007.02077}}].

\bibitem{DiVecchia:2021bdo}
P.~Di~Vecchia, C.~Heissenberg, R.~Russo, and G.~Veneziano, {\it {The eikonal
  approach to gravitational scattering and radiation at $ \mathcal{O}
  $(G$^{3}$)}},  {\em JHEP} {\bf 07} (2021) 169,
  [\href{http://arxiv.org/abs/2104.03256}{{\tt arXiv:2104.03256}}].

\bibitem{Saketh:2021sri}
M.~V.~S. Saketh, J.~Vines, J.~Steinhoff, and A.~Buonanno, {\it {Conservative
  and radiative dynamics in classical relativistic scattering and bound
  systems}},  \href{http://arxiv.org/abs/2109.05994}{{\tt arXiv:2109.05994}}.

\bibitem{Cachazo:2014fwa}
F.~Cachazo and A.~Strominger, {\it {Evidence for a New Soft Graviton Theorem}},
   \href{http://arxiv.org/abs/1404.4091}{{\tt arXiv:1404.4091}}.

\bibitem{Strominger:2017zoo}
A.~Strominger, {\it {Lectures on the Infrared Structure of Gravity and Gauge
  Theory}},  \href{http://arxiv.org/abs/1703.05448}{{\tt arXiv:1703.05448}}.

\bibitem{Braginsky1987}
V.~B. Braginsky and K.~S. Thorne, {\it Gravitational-wave bursts with memory
  and experimental prospects},  {\em Nature} {\bf 327} (1987), no.~6118
  123--125.

\bibitem{Baut-Lad}
Y.~F. Bautista and A.~Laddha, {\it {In preparation}}, .

\bibitem{Guevara:2018wpp}
A.~Guevara, A.~Ochirov, and J.~Vines, {\it {Scattering of Spinning Black Holes
  from Exponentiated Soft Factors}},  {\em JHEP} {\bf 09} (2019) 056,
  [\href{http://arxiv.org/abs/1812.06895}{{\tt arXiv:1812.06895}}].

\bibitem{Cheung:2018wkq}
C.~Cheung, I.~Z. Rothstein, and M.~P. Solon, {\it {From Scattering Amplitudes
  to Classical Potentials in the Post-Minkowskian Expansion}},  {\em Phys. Rev.
  Lett.} {\bf 121} (2018), no.~25 251101,
  [\href{http://arxiv.org/abs/1808.02489}{{\tt arXiv:1808.02489}}].

\bibitem{Bern:2019crd}
Z.~Bern, C.~Cheung, R.~Roiban, C.-H. Shen, M.~P. Solon, and M.~Zeng, {\it
  {Black Hole Binary Dynamics from the Double Copy and Effective Theory}},
  {\em JHEP} {\bf 10} (2019) 206, [\href{http://arxiv.org/abs/1908.01493}{{\tt
  arXiv:1908.01493}}].

\bibitem{Strominger:2014pwa}
A.~Strominger and A.~Zhiboedov, {\it {Gravitational Memory, BMS
  Supertranslations and Soft Theorems}},  {\em JHEP} {\bf 01} (2016) 086,
  [\href{http://arxiv.org/abs/1411.5745}{{\tt arXiv:1411.5745}}].

\bibitem{Laddha:2018vbn}
A.~Laddha and A.~Sen, {\it {Observational Signature of the Logarithmic Terms in
  the Soft Graviton Theorem}},  {\em Phys. Rev. D} {\bf 100} (2019), no.~2
  024009, [\href{http://arxiv.org/abs/1806.01872}{{\tt arXiv:1806.01872}}].

\bibitem{Sahoo:2021ctw}
B.~Sahoo and A.~Sen, {\it {Classical Soft Graviton Theorem Rewritten}},
  \href{http://arxiv.org/abs/2105.08739}{{\tt arXiv:2105.08739}}.

\bibitem{Pasterski:2015tva}
S.~Pasterski, A.~Strominger, and A.~Zhiboedov, {\it {New Gravitational
  Memories}},  {\em JHEP} {\bf 12} (2016) 053,
  [\href{http://arxiv.org/abs/1502.06120}{{\tt arXiv:1502.06120}}].

\bibitem{Bjerrum-Bohr:2021din}
N.~E.~J. Bjerrum-Bohr, P.~H. Damgaard, L.~Plant\'e, and P.~Vanhove, {\it {The
  Amplitude for Classical Gravitational Scattering at Third Post-Minkowskian
  Order}},  \href{http://arxiv.org/abs/2105.05218}{{\tt arXiv:2105.05218}}.

\bibitem{Kalin:2020fhe}
G.~K\"alin, Z.~Liu, and R.~A. Porto, {\it {Conservative Dynamics of Binary
  Systems to Third Post-Minkowskian Order from the Effective Field Theory
  Approach}},  {\em Phys. Rev. Lett.} {\bf 125} (2020), no.~26 261103,
  [\href{http://arxiv.org/abs/2007.04977}{{\tt arXiv:2007.04977}}].

\bibitem{Bern:2021dqo}
Z.~Bern, J.~Parra-Martinez, R.~Roiban, M.~S. Ruf, C.-H. Shen, M.~P. Solon, and
  M.~Zeng, {\it {Scattering Amplitudes and Conservative Binary Dynamics at
  ${\cal O}(G^4)$}},  {\em Phys. Rev. Lett.} {\bf 126} (2021), no.~17 171601,
  [\href{http://arxiv.org/abs/2101.07254}{{\tt arXiv:2101.07254}}].

\bibitem{Dlapa:2021npj}
C.~Dlapa, G.~K\"alin, Z.~Liu, and R.~A. Porto, {\it {Dynamics of Binary Systems
  to Fourth Post-Minkowskian Order from the Effective Field Theory Approach}},
  \href{http://arxiv.org/abs/2106.08276}{{\tt arXiv:2106.08276}}.

\bibitem{Sterman1993}
G.~Sterman, {\em An Introduction to Quantum Field Theory}.
\newblock Cambridge University Press, Aug., 1993.

\bibitem{Hannesdottir:2019opa}
H.~Hannesdottir and M.~D. Schwartz, {\it {$S$ -Matrix for massless particles}},
   {\em Phys. Rev. D} {\bf 101} (2020), no.~10 105001,
  [\href{http://arxiv.org/abs/1911.06821}{{\tt arXiv:1911.06821}}].

\bibitem{Kalin:2020mvi}
G.~K\"alin and R.~A. Porto, {\it {Post-Minkowskian Effective Field Theory for
  Conservative Binary Dynamics}},  {\em JHEP} {\bf 11} (2020) 106,
  [\href{http://arxiv.org/abs/2006.01184}{{\tt arXiv:2006.01184}}].

\end{thebibliography}\endgroup
\end{document}